\documentclass[twocolumn,english,showpacs,superscriptaddress]{revtex4-1}
\usepackage[T1]{fontenc}
\usepackage[latin9]{inputenc}
\usepackage{geometry}
\usepackage{xcolor}
\usepackage{babel}
\usepackage{units}
\usepackage{amsmath}
\usepackage{amssymb}
\usepackage{stmaryrd}
\usepackage{graphicx}
\usepackage{wasysym}
\usepackage{bbold}
\usepackage[bookmarks=false]{hyperref}

\makeatletter

\newcommand{\BL}[1]{{\color{black}#1}}

\IfFileExists{lmodern.sty}{\usepackage{lmodern}}{}
\usepackage{babel}

\usepackage{hyperref}

\makeatother

\begin{document}
\title{$N$-body interactions between trapped ion qubits via spin-dependent squeezing}
\date{\today}

\author{Or Katz}
\email{Corresponding author: or.katz@duke.edu}
\address{Duke Quantum Center, Duke University, Durham, NC 27701}
\address{Department of Electrical and Computer Engineering, Duke University, Durham, NC 27708}
\address{Department of Physics, Duke University, Durham, NC 27708}
\author{Marko Cetina}
\address{Duke Quantum Center, Duke University, Durham, NC 27701}
\address{Department of Physics, Duke University, Durham, NC 27708}

\author{Christopher Monroe}
\address{Duke Quantum Center, Duke University, Durham, NC 27701}
\address{Department of Electrical and Computer Engineering, Duke University, Durham, NC 27708}
\address{Department of Physics, Duke University, Durham, NC 27708}
\address{IonQ, Inc., College Park, MD  20740}

\begin{abstract}
We describe a simple protocol for the single-step generation of $N$-body entangling interactions between trapped atomic ion qubits. We show that qubit state-dependent squeezing operations and displacement forces on the collective atomic motion can generate full $N$-body interactions. Similar to the  M\o{}lmer-S\o{}rensen two-body Ising interaction at the core of most trapped ion quantum computers and simulators, the proposed operation is relatively insensitive to the state of motion. We show how this $N$-body gate operation allows the single-step implementation of a family of $N$-bit gate operations such as the powerful $N$-Toffoli gate, which flips a single qubit if and only if all other $N$-$1$ qubits are in a particular state.

\end{abstract}
\maketitle

The central ingredient in a quantum computer is the controllable quantum entanglement of its degrees of freedom, \BL{which for certain problems enables an exponential speed-up compared to classical algorithms.} The qubit and gate model of a quantum computer employs a universal set of operations, such as single-qubit rotations and two-qubit controlled-NOT gates \cite{Mike_and_Ike}. While such few-qubit interactions are sufficient for general computation, and can be used to construct many-body entangled states \cite{Sackett2000,monz2009realization,Monz2011,omran2019generation,Song2019,groenland2020signal}, many-qubit interactions can dramatically simplify quantum circuit structures, speed up their execution, and extend the power of quantum computer systems facing decoherence. For example, direct $N$-qubit operations such as the $N$-qubit Toffoli gate \cite{Toffoli1980} are expected to find native use in quantum adders and multipliers \cite{vedral1996quantum}, Grover searches \cite{grover1996fast,Wang2001,figgatt2017complete}, error-correction encoding \cite{paetznick2013universal,kitaev2003fault,muller2011simulating}, variational quantum algorithms for calculating electronic properties of molecules and materials \cite{o2016scalable,nam2020ground,seeley2012bravyi}, and simulations of nuclear structure and lattice gauge theories  \cite{banuls2020simulating,ciavarella2021trailhead,davoudi2020towards}.

Quantum gates based on many-body interactions have been proposed in several leading physical quantum platforms, from trapped ions \cite{CiracZoller1995,goto2004multiqubit,Espinoza2021,Wang2001} and neutral atoms \cite{weimer2010rydberg,isenhower2011multibit,Levine2019,khazali2020fast,xing2021realization} to superconducting systems \cite{Zahedinejad2015,khazali2020fast}. Here we concentrate on trapped ion qubits, which feature a high degree of control, very long qubit coherence times, high quantum gate fidelities and dense qubit connectivity \cite{wineland1998experimental,Brown2016,QEDC}. There have been proposals to realize $N$-body interactions between trapped ions using photons controlled by external optical cavities \cite{goto2004multiqubit} or phonons underlying the Coulomb-coupled motion of the ions \cite{CiracZoller1995,Espinoza2021,Wang2001}.
All of the above proposals rely on either having $N$ or more steps, special auxiliary qubit level structures, or exquisite control or pure-state preparation of the mediating phonons/photons.

Here we discover a simple single-step protocol for $N$-body entangling interactions between trapped ion qubits or effective spins. The operation is similar to the widely used M\o{}lmer-S\o{}rensen (MS) two-body interaction \cite{Molmer1999,Sorensen1999,Solano1999, Milburn2000}, which relies upon qubit state-dependent displacement forces.
We show instead, that state-dependent \textit{squeezing} forces can generate tunable $N$-body interactions between the qubits. \BL{Such squeezing operations have previously been considered in other proposals \cite{drechsler2020state,sutherland2021motional,bohnet2016quantum,feng2021sensitive}, and the context of gates, spin-independent motional squeezing has been used to enhance MS gate performance, but none of these works change the form of the underlying two-body spin interaction \cite{ge2019trapped,ge2019stroboscopic,burd2021quantum}.}
We further present a limiting case of the protocol and demonstrate the construction of the $N$-Toffoli gate in a single step, and discuss other classes of $N$-body spin Hamiltonians that can be similarly generated.  As in the MS protocol \cite{Molmer1999, Sorensen1999,Solano1999, Milburn2000,Wang2001}, this scheme does not rely on a pure initial phonon state and can be relatively insensitive to thermal motion.

\begin{figure*}[t]
\begin{centering}
\includegraphics[width=17.7cm]{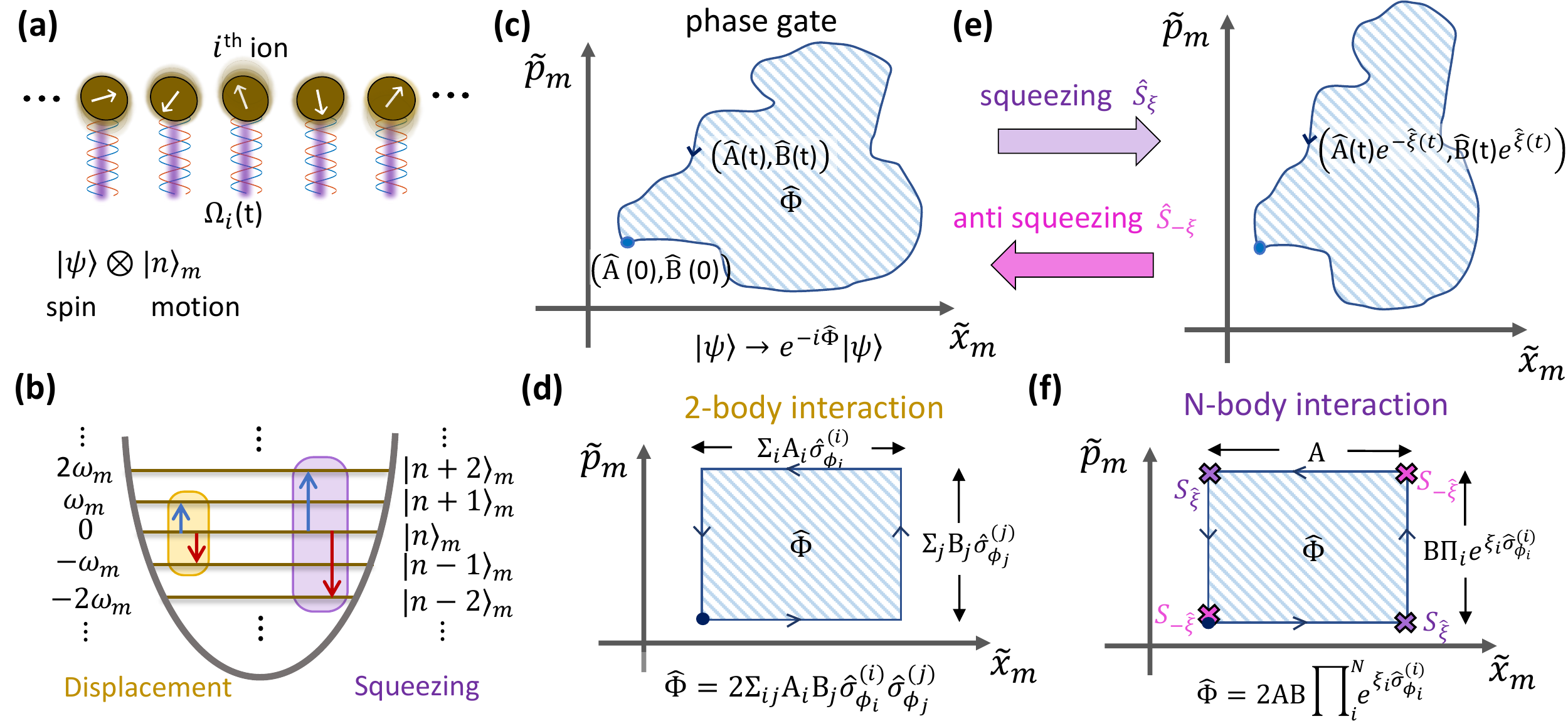}
\par\end{centering}
\centering{}\caption{Quantum Phase Gates with trapped ions. (a) A chain of trapped ions whose many-body spin state $|\psi \rangle$ is decoupled from the motional state $|n\rangle_m$ of a single harmonic phonon mode $m$ represented by vibrational integer index $n \ge 0$. Ions are addressed with bi-chromatic Laser fields with carrier spin-flip Rabi rates $\Omega_i$. (b) Motional sideband transitions driven by the laser field. Tuning the laser field on resonance with the first red and blue sideband transitions at frequency $\pm\omega_m$ from the carrier \cite{Leibfried2003} generates a spin-dependent force through the absorption and \BL{of} emission phonons. Tuning the tones at the second red and blue sidebands at $\pm 2\omega_m$ from the carrier generates spin-dependent squeezing by absorption and emission of pairs of phonons. (c) Displacing the motion of mode $m$ in a closed loop of phase space adds a phase $\hat{\Phi}$ to the quantum state that is given by the area of the enclosed contour. (d) The MS gate using spin-dependent displacements result in a spin-dependent phase linear in the \BL{spin operators $\hat{\sigma}_{\phi_i}^{\left(i\right)}\equiv\hat{\sigma}_{x}^{\left(i\right)}\cos\phi_{i}+\hat{\sigma}_{y}^{\left(i\right)}\sin\phi_{i}$ of ion $i$}. When applied to multiple ions, the resulting phase $\hat{\Phi}$ is thus quadratic in the spin operators, corresponding to 2-body MS interaction \cite{sorensen2000entanglement,Milburn2000}. (e) Motional-squeezing shrinks one direction in phase space but expands the other to conserve the phase space element area. (f) $N$-body entangling gate. Synchronized spin-dependent squeezing (cross symbols) applied in between displacements produces squeezing of the motion along momentum axis. The phase $\hat{\Phi}$ now depends exponentially on the spins, and therefore contains products of N spin operators. The phase space axes are displayed with dimensionless units ${\tilde{x}}_m={\hat{x}}_m/(2x^0_m)$ and ${\tilde{p}}_m={\hat{p}}_m/(2p^0_m)$, with the convention $[{\tilde{x}}_m,{\tilde{p}}_m]=i/2$.
\label{fig:Fig1}}
\end{figure*}

The central idea behind trapped ion quantum gates is the coupling between spins and motion (phonons) through spin-dependent forces \cite{CiracZoller1995,Molmer1999,Sorensen1999,Solano1999, Milburn2000}, as illustrated in Fig.~\ref{fig:Fig1}. Owing to the Coulomb interaction between the trapped ions, their motion around equilibrium can be expressed by collective normal modes of harmonic oscillation.
We focus on the coupling through a single phonon mode through a near-resonance driving force, although generalization to multiple modes is straightforward \cite{Zhu2006,Debnath2016}. 
We represent the phonon state of mode $m$ in a frame that rotates at the mode oscillation frequency $\omega_m$ using the phase-space position and momentum operators $\hat{x}_m=x^0_{m}(\hat{a}_m^{\dagger}+\hat{a}_m),\,\hat{p}_m=ip^0_{m}(\hat{a}_m^{\dagger}-\hat{a}_m)$. Here, $\hat{a}_m^{\dagger} (\hat{a}_m)$ are the bosonic creation (annihilation) operators and $x^0_{m}=\sqrt{\hbar/2M\omega_m}$ ($p^0_{m}=\sqrt{\hbar M\omega_m/2})$ are the zero-point spread in position (momentum) associated with mode $m$, where $M$ is the mass of a single ion. The spin-motion coupling is parametrized by the Lamb-Dicke parameter $\eta_{im}=b_{im}\eta_m$, where $\eta_m=kx^0_{m}$, $k$ is the effective wavenumber of the field driving the motion \cite{Leibfried2003} and $b_{im}$ is the mode participation matrix between ion $i$ and mode $m$, with $\sum_i b_{im}b_{in} = \delta_{nm}$ and $\sum_m b_{im}b_{jm} = \delta_{ij}$.

The MS interaction arises by addressing multiple ions on the first red and blue sidebands of mode $m$ from the spin-flip carrier, with relative phase $\delta\phi_i$ and zero-point Rabi rate $\eta_{im}\Omega_i(t)$ for ion $i$. The carrier Rabi frequency $\Omega_i(t)$ is proportional to the drive strength, and we assume the motion is confined within the Lamb-Dicke regime $(\eta_{im}\langle \hat{a}_m^{\dagger}+\hat{a}_m \rangle \ll 1)$ \cite{Leibfried2003}. 
This spin-dependent force displaces the phonon state in phase space by position $\hat{\mathrm{A}}(t)=\sum_i A_i(t)\hat{\sigma}_{x}^{(i)}$ and momentum $\hat{\mathrm{B}}(t)=\sum_i B_i(t)\hat{\sigma}_{x}^{(i)}$, where $\hat{\sigma}_{x}^{(i)}$ are the Pauli spin flip operators (chosen uniformly along $x$ for convenience). The position and momentum displacement amplitudes, scaled by $2x_m^0$ and $2p_m^0$, are $A_{i}(t)=\frac{1}{2}\eta_{im}\int_0^tdt'\Omega_i\sin\delta\phi_i$ and $B_{i}(t)=\frac{1}{2}\eta_{im}\int_0^t dt'\Omega_i\cos\delta\phi_i$ \cite{Molmer1999}.

Geometric phase gates such as the MS gate displace the ions in closed phase space loops [Fig.~\ref{fig:Fig1}(c) and (d)]. By the end of the gate at time $T$, the spin state of the ions is decoupled from the phonons but has evolved according to $U_{\textrm{MS}}(T)=e^{-i\hat{\Phi}}$, with geometrical phase operator 
\begin{equation}\label{eq:Phi_def}
\hat{\Phi}=-2\int_{0}^{T}\hat{\mathrm{B}}(t)\frac{d\hat{\mathrm{A}}(t)}{dt}dt.
\end{equation}
Because $\hat{\mathrm{A}}(t)$ and $\hat{\mathrm{B}}(t)$ are linear in the spin operators, the gate phase operator $\hat{\Phi}$ is quadratic in the spin operators \cite{lu2019global,figgatt2019parallel}, limiting the standard MS gate to two-body (Ising) interactions.

To generate an $N$-body spin interaction, we consider the effect of spin-dependent motional squeezing on a phase gate operation. Spin-dependent squeezing can be generated by driving the \textit{second} red and blue sidebands of a single phonon mode $m$, as shown in Fig.~\ref{fig:Fig1}(b) \cite{Leibfried2003}. Setting the zero-point 2\textsuperscript{nd} sideband Rabi rates equal to $\eta_{im}^2\Omega_{i}/2$ and the relative phase between the drives constant across the chain ($\delta\phi_{i}=\delta\phi$), the spin-motion interaction becomes, \begin{equation}
H_{S}=\frac{i\hbar}{4}\Bigl(e^{i\delta\phi\BL{(t)}}\hat{a}_m^{2}-e^{-i\delta\phi\BL{(t)}}\hat{a}_m^{\dagger2}\Bigr)\sum_{i=1}^{N}\eta_{im}^{2}\Omega_{i}\BL{(t)}\hat{\sigma}_{\phi_i}^{(i)}.\label{eq:spin-dependent squeezing}
\end{equation}
Here 
$\hat{\sigma}_{\phi_i}^{\left(i\right)}\equiv\hat{\sigma}_{x}^{\left(i\right)}\cos\phi_{i}+\hat{\sigma}_{y}^{\left(i\right)}\sin\phi_{i}$ 
is the Pauli spin-flip operator of spin $i$ set by the average phase $\phi_{i}$ of the $i^{\textrm{th}}$ pair of drives. The phonon operator in Eq.~(\ref{eq:spin-dependent squeezing}) is the generator of quadrature squeezing along the axis rotated by $\delta\phi/2$ and anti-squeezing along $(\pi+\delta\phi)/2$; we fix $\delta\phi(t)=0$ for convenience.
\begin{figure}
\begin{centering}
\includegraphics[width=8.4cm]{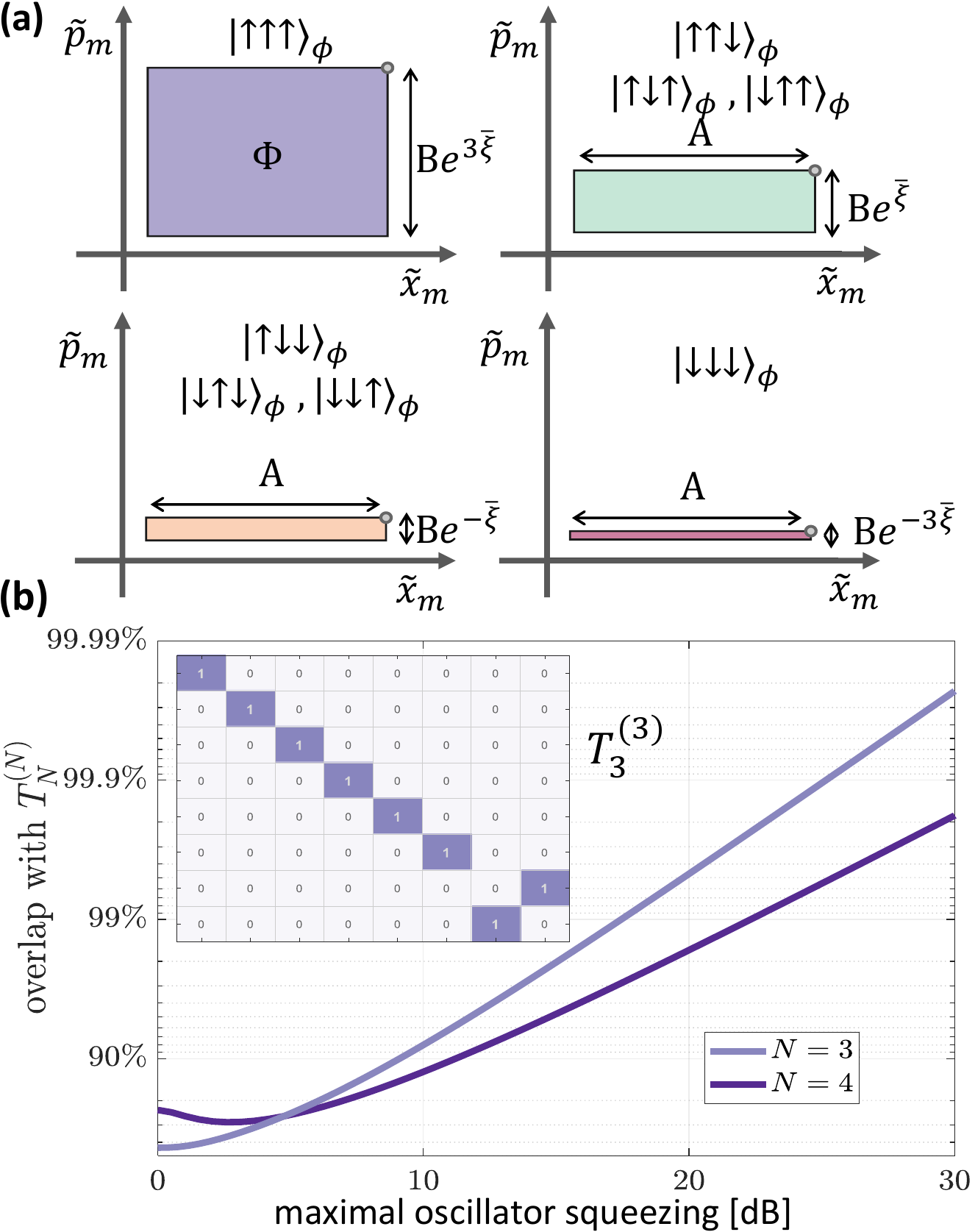}
\par\end{centering}
\centering{}\caption{3-body entangling gates. (a) Phase space evolution for three spins, following the sequence of alternating spin-independent displacements and spin-dependent squeezing operations [c.f.~Fig.~\ref{fig:Fig1}(f) and Eq.~(\ref{eq:seq_operations})]. Each ion squeezes (anti-squeezes) the momentum quadrature of the $m$th motional mode by a factor $e^{-\bar\xi}$ ($e^{\bar\xi}$) if its spin points downwards (upwards). The state-dependent phase-space area $\hat{\Phi}_{\mathrm{seq}}$ accumulated in the evolution generates the gate  $U_{\mathrm{seq}}=e^{-i\hat{\Phi}_{\mathrm{seq}}}$ with a maximal squeezing of the oscillator mode by a factor $e^{N\bar{\xi}}$ when all spins are aligned.
(b) Overlap between the proposed many-body gate in Eq.~(\ref{eq:phase}) 
(accompanied by single-qubit rotations on the third qubit as described in the main text) and the $N$-Toffoli gate $T_{N}^{(N)}$ depending on the maximal squeezing of phase-space coordinates in dB ($10\log_{10}{(e^{N\bar{\xi}})}$) for $N=3,4$. Inset: ideal $T_{3}^{(3)}$ operator in the computational basis.
\label{fig:Fig2}}
\end{figure}
Under the time-dependent Hamiltonian $H_{S}$ of Eq.~(\ref{eq:spin-dependent squeezing}), the quantum state evolves by the spin-dependent squeezing operator \BL{\cite{drechsler2020state}}
\begin{equation}
S_{\hat{\xi}}(t)=e^{\frac{1}{2}\hat{\xi}(t)(\hat{a}_m^{2}-\hat{a}_m^{\dagger2})}, 
\end{equation}
where the spin-dependent squeezing amplitude is
\begin{equation}\label{eq:xi}
\hat{\xi}(t)=\sum_{i}\xi_i(t)\hat{\sigma}_{\phi_i}^{\left(i\right)}=\frac{1}{2}\sum_{i}\hat{\sigma}_{\phi_i}^{\left(i\right)}\eta_{im}
^2\int_0^t\Omega_i(\tau)d\tau.
\end{equation}

To illustrate the effect of squeezing on a phase gate operation, we first consider an alternating sequence of spin-dependent squeezing operations and displacement forces.
Specifically, we apply four discrete displacements along a rectangular-shaped closed-loop in phase space \cite{Milburn2000} interspersed with four alternating squeezing operators applied at the corners of the rectangle that ultimately decouple the motion, as depicted in Fig.~\ref{fig:Fig1}(f).
The displacements in position $\hat{\mathrm{A}}(t_x)$ and momentum $\hat{\mathrm{B}}(t_p)$ are applied for times $t_x$ and $t_p$, respectively. Each squeezing operators is applied for time $t_S$ with squeezing amplitude $\hat{\xi}(t_S)$, for a total gate time of $T=4t_S+2t_x+2t_p$. The evolution operator of this sequence is written


\begin{eqnarray}
U_{\textrm{seq}}(T) &=& S_{\hat{\xi}}^{\dagger}D(-i{\hat{\mathrm{B}}})S_{\hat{\xi}}D(-{\hat{\mathrm{A}}})S_{\hat{\xi}}^{\dagger}D(i{\hat{\mathrm{B}}})S_{\hat{\xi}}D({\hat{\mathrm{A}}})\label{eq:seq_operations} \\
&=& D(-i{\hat{\mathrm{B}}}e^{\hat{\xi}})D(-{\hat{\mathrm{A}}})D(i{\hat{\mathrm{B}}}e^{\hat{\xi}})D(\hat{\mathrm{A}}) \label{eq:contracted} \\ &=&e^{-i\hat{\Phi}_{\textrm{seq}}} \label{phasegate},
\end{eqnarray}
where $D(\alpha )=e^{\alpha \hat{a}^{\dagger}_m- \alpha^{*} \hat{a}_m}$ is the displacement operator, which moves the phonon state in phase space by $2x^0_m{\textrm{Re}(\alpha)}$ along the $\hat{x}_m$ coordinate and by $2p^0_m{\textrm{Im}(\alpha)}$ along $\hat{p}_m$.
The squeezing operations produce a net displacement whose magnitude is 
dilated or contracted depending on the spin, since $S_{\hat{\xi}}^{\dagger}D(i{\hat{\mathrm{B}}})S_{\hat{\xi}}\equiv D(i{\hat{\mathrm{B}}}e^{\hat{\xi}})$. Because $\hat{\xi}$ is linear in the spin operators from Eq.~(\ref{eq:xi}), the 
gate phase operator is \textit{exponential} in the spin operators: \begin{equation}\label{eq:phase}
\hat{\Phi}_{\textrm{seq}}=2\hat{\mathrm{A}}\hat{\mathrm{B}}e^{\hat{\xi}}
=2\hat{\mathrm{A}}\hat{\mathrm{B}}\prod_{i=1}^{N}\left(\mathbb{1}\cosh\xi_{i}+\hat{\sigma}_{\phi_i}^{\left(i\right)}\sinh\xi_{i}\right),
\end{equation}
corresponding to an effective $N$-body Hamiltonian
$H_{\mathrm{eff}}=\hbar\hat{\Phi}_{\textrm{seq}}/T$. This remarkable construction features many-body interaction terms, where the relative contribution of the $N$-body term scales as $\prod_i \tanh{\xi_i}$, which is sizeable for \BL{$\xi_i\sim 1.$}

\BL{We now demonstrate the protocol with a few simple gates that can be cast in the form of Eq.~(\ref{eq:phase}). First, we consider a three-body gate between qubits $i,j,k$ given by \begin{equation} \label{eq:three_qubit_gate}
U_{\textrm{seq}}^{(ijk)}=\exp{(-i\varphi\hat{\sigma}_{\phi_i}^{\left(i\right)}\hat{\sigma}_{\phi_j}^{\left(j\right)}\hat{\sigma}_{\phi_k}^{\left(k\right)})},
\end{equation}
that can find usage in various applications, e.g.~\cite{marvian2022restrictions,roggero2020quantum,pachos2004three}. This gate can be realized via spin-dependent displacements of the $i,j$ spins generating $\hat{\mathrm{A}}=\mathrm{A}_i\hat{\sigma}_{\phi_i}^{\left(i\right)}$ and $\hat{\mathrm{B}}=\mathrm{B}_j\hat{\sigma}_{\phi_j}^{\left(j\right)}$ and squeezing by spin $k$, generating $\hat{\xi}=\xi_{k}\hat{\sigma}_{\phi_k}^{\left(k\right)}$. Eq.~(\ref{eq:three_qubit_gate}) is then obtained for $\mathrm{A}_i\mathrm{B}_j\cosh(\xi_k)=\pi/2$ and $\varphi=\pi\tanh{\xi_k}$. Similar to MS gates \cite{sorensen2000entanglement}, maximally entangled states can be prepared for $\varphi=\pi/4$ which remarkably correspond to squeezing of the oscillator mode by about $\xi_k\approx0.25$ which is $10\log_{10}(e^{\xi_k})\approx 1\,\mathrm{dB}$.

As a second example, we consider simultaneous squeezing of $N$ spins but spin-independent displacements $\hat{\mathrm{A}}(t_x)=\mathrm{A}\mathbb{1}$ and  $\hat{\mathrm{B}}(t_p)=\mathrm{B}\mathbb{1}$ where $\mathrm{A}=\sum_i A_i(t), \mathrm{B}=\sum_i B_i(t)$ and $\mathbb{1}$ is the identity spin operator. We plot the phase-space trajectories of this configuration in Fig.~\ref{fig:Fig2}a for $N=3$ qubits, assuming a common squeezing amplitude $\xi_i=\bar{\xi}$. The phase accumulated by the quantum state depends exponentially on the number of spins pointing upward. In the limit ${\xi}_i\gg1$, the phase operator in Eq.~(\ref{eq:phase}) becomes}
\begin{equation} \label{eq:proj}
\hat{\Phi}_{\textrm{seq}}\rightarrow 2\mathrm{A}\mathrm{B}e^{N\bar{\xi}}\prod_{i=1}^{N}\tfrac{1}{2}\left(\mathbb{1}+\hat{\sigma}_{\phi_i}^{\left(i\right)}\right).
\end{equation}
Eq.~(\ref{eq:proj}) is proportional to the projection operator on the state ${\mid\uparrow_{\phi}\cdots\uparrow_{\phi}\rangle}$, in which each spin points upward along an eigenstate of $\hat{\sigma}_{\phi_i}^{(i)}$.
This renders $U_{\textrm{seq}}$ into the $N$-qubit controlled-phase gate, which appends the phase factor $\mathrm{exp}(-2i\mathrm{A}\mathrm{B}e^{N\bar{\xi}})$ to the state ${\mid\uparrow_{\phi}\cdots\uparrow_{\phi}\rangle}$. From here, it is easy to construct the $N$-bit Toffoli gate $T_{N}^{(n)}$, which flips qubit $n$ if and only if all other $N-1$ qubits point up \cite{Toffoli1980,CiracZoller1995}. By setting $2\mathrm{A}\mathrm{B}e^{N\bar{\xi}}=\pi$ and surrounding this operation by single-qubit $\pi/2$ rotations on qubit $n$, we find $T_{N}^{(n)}=R^{(n)}_z(\pi/2)U_{\textrm{seq}}R^{(n)}_z(-\pi/2)$. 

\BL{To characterize the action of the proposed unitary gate $\tilde{U}_{\textrm{seq}}$ at finite squeezing amplitudes, we calculate its overlap \footnote{We calculate the overlap as the entanglement fidelity of the two unitary processes. c.f.~\cite{horodecki1999general}.} with the ideal Toffoli gate as a function of the maximal degree of squeezing of the oscillator mode, shown in Fig.~\ref{fig:Fig2}b for $N=3,4$. We find that the overlap approaches unity at high levels of squeezing. Notably, nonideal overlap does not imply nonunitary evolution or an error, but rather that the amplitudes of the spin terms in the actual unitary gate [Eq.~(\ref{eq:phase})] are not exactly all equal as in the ideal Toffoli [Eq.~(\ref{eq:proj})]. For some applications, e.g.~variational quantum algorithms \cite{zhou2020quantum,cerezo2021variational,kandala2017hardware}, it can usefully expand the native set of gates even with moderate overlap.}


We next generalize the sequential protocol in Eq.~(\ref{eq:seq_operations}) and consider simultaneous application of displacements and spin-dependent squeezing. Displacements are generated in the interaction picture by the Hamiltonian \begin{equation}
H_{D}=2x^0_m\hat{\alpha}(t)\hat{p}_m - 2p^0_m\hat{\beta}(t) \hat{x}_m.
\label{eq:spin_independent_displacement}
\end{equation}
where the forces $\hat{\alpha},\hat{\beta}$ are Hermitian, and their spin-dependence is determined by the underlying mechanism from which they are produced. For example, fields produced by the trap's electrodes couple to the ions' charge and can generate state-independent displacements \cite{wineland1998experimental}, whereas optical dipole forces \cite{Monroe2021,gilmore2021quantum,haljan2005spin} or magnetic fields \cite{mintert2001ion,harty2016high,srinivas2021high} can generate displacements linear in the spin operators. 

The total Hamiltonian of the system is then given by $H(t)=H_{S}(t)+H_{D}(t)$, and the time-ordered evolution operator can be represented by \cite{wei1963lie} \begin{equation}\label{eq:U_tot}U(t)=S_{\hat{\xi}}(t)U_{D}(t).\end{equation} The operator $U_{D}$ describes the contribution of phase-space displacements to the evolution and is generated by the Hamiltonian
\begin{equation}
{H}'_{D}=S^{\dagger}_{\hat{\xi}}H_{D} S_{\hat{\xi}}
=2x^0_m\hat{\alpha}(t) e^{\hat{\xi}(t)}\hat{p}_m - 2p^0_m\hat{\beta}(t) e^{-\hat{\xi}(t)}\hat{x}_m,
\label{eq:spin_dependent_displacement}
\end{equation}
provided that $\hat{\alpha},\hat{\beta}$ and $\hat{\xi}$ commute during the gate. Spin-dependent squeezing thus renders the standard forces $\hat{\alpha},\hat{\beta}$ as nonlinear in the spin operators, via the exponential terms $e^{\pm\hat{\xi}(t)}$ in Eq.~(\ref{eq:spin_dependent_displacement}). Yet, the evolution of $U_D$ is identical to that of the MS gate under the simple transformation $\hat{\alpha}\rightarrow\hat{\alpha}e^{\hat{\xi}},$  $\hat{\beta}\rightarrow\hat{\beta}e^{-\hat{\xi}}$ and is therefore described by \cite{sorensen2000entanglement} \begin{equation}
U_D(t)=e^{-i\hat{\Phi}}D\bigl(i\hat{\mathrm{B}}\bigr)D\bigl(\hat{\mathrm{A}}\bigr). \label{eq:Unitary-evolution}\end{equation}
The Hermitian phase-space displacements are given by \begin{equation}\begin{aligned}\hat{\mathrm{A}}(t)&=\int_{0}^{t}e^{\hat{\xi}(t')}\hat{\alpha}(t')dt',\\\hat{\mathrm{B}}(t)&=\int_{0}^{t}e^{-\hat{\xi}(t')}\hat{\beta}(t')dt',\end{aligned}\end{equation}
and the phase operator $\hat{\Phi}(T)$ by Eq.~(\ref{eq:Phi_def}).
Similar to the MS gate, the operator $U$ in Eq.~(\ref{eq:U_tot}) entangles the spin and motional states during the gate operation. To realize a gate that is independent of motion for all input states, we require that at $t=T$,
\begin{equation}\label{eq:closure_conditions}
\hat{\mathrm{A}}(T)=\hat{\mathrm{B}}(T)=\hat{\xi}(T)=0.
\end{equation} 
This decouples the motion (both in displacement and squeezing) so that the net evolution operator contains only spin operators, yielding $U(T)=e^{-i\hat{\Phi}}$. 
\BL{In \cite{SI} we present several examples of effective Hamiltonians that are generated by simultaneous application of displacement and squeezing operations as well as a protocol to generate the 4-body gate  $\exp{\bigl(-i\tfrac{\pi}{4}\hat{\sigma}_{\phi_i}^{\left(i\right)}\hat{\sigma}_{\phi_j}^{\left(j\right)}\hat{\sigma}_{\phi_k}^{\left(k\right)}\hat{\sigma}_{\phi_l}^{\left(l\right)}}\bigr)$, which can be used to simulate e.g.~the plaquette operators in lattice-gauge theories or the Toric-code Hamiltonian \cite{davoudi2020towards,kitaev2003fault}.}

We now consider the speed of the $N$-body gate, especially as it relies on 2\textsuperscript{nd} order motional sidebands, which in the Lamb-Dicke limit are weak. 
For two-body MS gates through single mode $m$ \cite{sorensen2000entanglement}, the gate time can be as short as $T\approx \pi/(b_{im}\eta_m\Omega)$,
taking a uniform Rabi frequency $\Omega$ over the involved ions.
The $N$-body gate presented here is based on resonant spin-dependent squeezing \BL{operations with additional duration of }$4t_S\approx 8\bar{\xi}/(\Omega b_{im}^2\eta_m^2)$. \BL{We estimate these parameters for a chain of 4 Yb ions in a trap with radial frequency of $2\pi\times3\,\mathrm{MHz}$ and $\eta_m\approx0.11$, corresponding to coherent driving by counter-propagating $355\,\mathrm{nm}$ light. The action of this light on the 2\textsuperscript{nd} sideband with a typical value of $\Omega=2\pi\times1\,\mathrm{MHz}$ squeezes the oscillator mode at the rate of about $d\xi/dt\approx 10\,\mathrm{ms}^{-1}$ per ion (which corresponds to $40\,\mathrm{dB/ms}$), for a typical mode participation factor of $b_{im}=0.5$. We focus on low values of $\xi$ as highly squeezed physical systems may become more susceptible to experimental imperfections. For the fully-entangling three qubit gate in Eq.~(\ref{eq:three_qubit_gate}) with $\varphi=\pi/4,$ we estimate a typical squeezing time of $t_S\approx25\,\mu\mathrm{s}$, and a total gate time of about $T\approx 130\,\mu\mathrm{s}$. 
This duration is similar to that of other practically realized gates \cite{egan2021fault,wang2020high,ballance2016high}, and is considerably shorter than the motional coherence time of the oscillator (typically $10-100$ ms \cite{wang2020high,burd2021quantum}), which is expected to be the limiting factor for the gate performance. Using low-mass $^9$Be$^+$ ions with $\eta\approx0.25$ \cite{gaebler2016high} is expected to shorten the gate time about $5$-fold. For longer chains, scaling of the mode participation factors by $b_{im}\sim1/\sqrt{N}$ yields a linear scaling of the gate duration in $N$, although the gates can be tailored to be faster for certain subsets of ions by using more localized modes.}

Our analysis focuses on the interactions generated via resonant coupling with a single phonon mode. However, spin-dependent squeezing through 2\textsuperscript{nd} sidebands can also drive off-resonant sidebands on pairs of modes $\mu,\nu$ detuned by $\Delta_{\mu\nu}=2\omega_m-\omega_{\mu}-\omega_{\nu}$.
This results in multi-mode squeezing in a potentially dense sideband spectrum, with the possibility of near-degeneracies. These off-resonant couplings can be suppressed by judiciously shaping the axial ion trap potential and choosing the target mode so that the unwanted sidebands are sufficiently far from the desired squeezing sideband. 
For example, the lowest frequency (zig-zag) radial normal mode is relatively isolated, and 
the resulting off-resonant coupling with the nearest 2\textsuperscript{nd} sideband detuned $\Delta_{\mu m}$ from the drive scales with rate $\frac{1}{4}\sum_i\eta_{i\mu}^2\eta_{im}^2\Omega_i^2/\Delta_{\mu m}$, while the desired squeezing interaction rate scales as $\eta_m^2\Omega/(2N)$, \BL{the ratio reads as $\epsilon = \eta_m^2\Omega/(2N\Delta_{\mu m})$ where excitation of phonons scale as $\epsilon^2$. For the 4 ions configuration with a lowest axial frequency of $900\,\textrm{kHz}$, we find $\Delta\approx160\,\textrm{kHz}$ for the lowest frequency mode and $\epsilon^2\lesssim0.01$}. By shaping the mode spectrum such that $\Delta_{\mu m} \sim \mathcal{B}/N$ for instance, where $\mathcal{B}$ is the bandwidth of modes, we find that for fixed $\Omega$, $\epsilon$ does not grow with $N$.
Furthermore, it is possible to apply pulse-shaping techniques to control all multimode squeezing operations for the $N$-body gate while decoupling all motional modes, exactly as has been demonstrated for multimode MS gates \cite{Zhu2006,Debnath2016}.

We finally note that the emergence of $N$-body interactions discovered here can be seen from the expanded Lie algebra generated by the combined squeezing and displacement Hamiltonians. This is evident from the Magnus expansion representation of the evolution operator \cite{Magnus1954}, a sequence of nested commutators of the Hamiltonian with itself. For the MS interaction, the series terminates after the second term because $[\hat{x}_m,\hat{p}_m]=i\hbar$. Here instead, the series does not terminate because for instance
$[(\hat{a}_m^2-\hat{a}_m^{\dagger2})\hat{\sigma_x}^{(i)},\hat{x}_m]=2\hat{x}_m\hat{\sigma_x}^{(i)}$, thus carrying products of further spin operators along in the expansion.
This interaction thus represents a new degree of freedom in controlling trapped ion quantum states, and may significantly expand the expression of trapped ion quantum logic operations.
\begin{acknowledgments}
This work is supported by the ARO through the IARPA LogiQ program; the NSF STAQ program; the DOE QSA program; the AFOSR MURIs on Dissipation Engineering in Open Quantum Systems, Quantum Measurement/Verification, and Quantum Interactive Protocols; the ARO MURI on Modular Quantum Circuits; and by the U.S. Department of Energy HEP QuantISED Program through the GeoFlow Grant No. de-sc0019380.
\end{acknowledgments}

\bibliography{Refs}

\begin{thebibliography}{71}%
\makeatletter
\providecommand \@ifxundefined [1]{%
 \@ifx{#1\undefined}
}%
\providecommand \@ifnum [1]{%
 \ifnum #1\expandafter \@firstoftwo
 \else \expandafter \@secondoftwo
 \fi
}%
\providecommand \@ifx [1]{%
 \ifx #1\expandafter \@firstoftwo
 \else \expandafter \@secondoftwo
 \fi
}%
\providecommand \natexlab [1]{#1}%
\providecommand \enquote  [1]{``#1''}%
\providecommand \bibnamefont  [1]{#1}%
\providecommand \bibfnamefont [1]{#1}%
\providecommand \citenamefont [1]{#1}%
\providecommand \href@noop [0]{\@secondoftwo}%
\providecommand \href [0]{\begingroup \@sanitize@url \@href}%
\providecommand \@href[1]{\@@startlink{#1}\@@href}%
\providecommand \@@href[1]{\endgroup#1\@@endlink}%
\providecommand \@sanitize@url [0]{\catcode `\\12\catcode `\$12\catcode
  `\&12\catcode `\#12\catcode `\^12\catcode `\_12\catcode `\%12\relax}%
\providecommand \@@startlink[1]{}%
\providecommand \@@endlink[0]{}%
\providecommand \url  [0]{\begingroup\@sanitize@url \@url }%
\providecommand \@url [1]{\endgroup\@href {#1}{\urlprefix }}%
\providecommand \urlprefix  [0]{URL }%
\providecommand \Eprint [0]{\href }%
\providecommand \doibase [0]{https://doi.org/}%
\providecommand \selectlanguage [0]{\@gobble}%
\providecommand \bibinfo  [0]{\@secondoftwo}%
\providecommand \bibfield  [0]{\@secondoftwo}%
\providecommand \translation [1]{[#1]}%
\providecommand \BibitemOpen [0]{}%
\providecommand \bibitemStop [0]{}%
\providecommand \bibitemNoStop [0]{.\EOS\space}%
\providecommand \EOS [0]{\spacefactor3000\relax}%
\providecommand \BibitemShut  [1]{\csname bibitem#1\endcsname}%
\let\auto@bib@innerbib\@empty
\bibitem [{\citenamefont {Nielsen}\ and\ \citenamefont
  {Chuang}(2000)}]{Mike_and_Ike}%
  \BibitemOpen
  \bibfield  {author} {\bibinfo {author} {\bibfnamefont {M.~A.}\ \bibnamefont
  {Nielsen}}\ and\ \bibinfo {author} {\bibfnamefont {I.~L.}\ \bibnamefont
  {Chuang}},\ }\href@noop {} {\emph {\bibinfo {title} {Quantum Computation and
  Quantum Information}}}\ (\bibinfo  {publisher} {Cambridge University Press},\
  \bibinfo {address} {Cambridge, UK},\ \bibinfo {year} {2000})\BibitemShut
  {NoStop}%
\bibitem [{\citenamefont {Sackett}\ \emph {et~al.}(2000)\citenamefont
  {Sackett}, \citenamefont {Kielpinski}, \citenamefont {King}, \citenamefont
  {Langer}, \citenamefont {Meyer}, \citenamefont {Myatt}, \citenamefont {Rowe},
  \citenamefont {Turchette}, \citenamefont {Itano}, \citenamefont {Wineland},\
  and\ \citenamefont {Monroe}}]{Sackett2000}%
  \BibitemOpen
  \bibfield  {author} {\bibinfo {author} {\bibfnamefont {C.~A.}\ \bibnamefont
  {Sackett}}, \bibinfo {author} {\bibfnamefont {D.}~\bibnamefont {Kielpinski}},
  \bibinfo {author} {\bibfnamefont {B.~E.}\ \bibnamefont {King}}, \bibinfo
  {author} {\bibfnamefont {C.}~\bibnamefont {Langer}}, \bibinfo {author}
  {\bibfnamefont {V.}~\bibnamefont {Meyer}}, \bibinfo {author} {\bibfnamefont
  {C.~J.}\ \bibnamefont {Myatt}}, \bibinfo {author} {\bibfnamefont
  {M.}~\bibnamefont {Rowe}}, \bibinfo {author} {\bibfnamefont {Q.~A.}\
  \bibnamefont {Turchette}}, \bibinfo {author} {\bibfnamefont {W.~M.}\
  \bibnamefont {Itano}}, \bibinfo {author} {\bibfnamefont {D.~J.}\ \bibnamefont
  {Wineland}},\ and\ \bibinfo {author} {\bibfnamefont {C.}~\bibnamefont
  {Monroe}},\ }\href {https://doi.org/10.1038/35005011} {\bibfield  {journal}
  {\bibinfo  {journal} {Nature}\ }\textbf {\bibinfo {volume} {404}},\ \bibinfo
  {pages} {256} (\bibinfo {year} {2000})}\BibitemShut {NoStop}%
\bibitem [{\citenamefont {Monz}\ \emph {et~al.}(2009)\citenamefont {Monz},
  \citenamefont {Kim}, \citenamefont {H{\"a}nsel}, \citenamefont {Riebe},
  \citenamefont {Villar}, \citenamefont {Schindler}, \citenamefont {Chwalla},
  \citenamefont {Hennrich},\ and\ \citenamefont {Blatt}}]{monz2009realization}%
  \BibitemOpen
  \bibfield  {author} {\bibinfo {author} {\bibfnamefont {T.}~\bibnamefont
  {Monz}}, \bibinfo {author} {\bibfnamefont {K.}~\bibnamefont {Kim}}, \bibinfo
  {author} {\bibfnamefont {W.}~\bibnamefont {H{\"a}nsel}}, \bibinfo {author}
  {\bibfnamefont {M.}~\bibnamefont {Riebe}}, \bibinfo {author} {\bibfnamefont
  {A.}~\bibnamefont {Villar}}, \bibinfo {author} {\bibfnamefont
  {P.}~\bibnamefont {Schindler}}, \bibinfo {author} {\bibfnamefont
  {M.}~\bibnamefont {Chwalla}}, \bibinfo {author} {\bibfnamefont
  {M.}~\bibnamefont {Hennrich}},\ and\ \bibinfo {author} {\bibfnamefont
  {R.}~\bibnamefont {Blatt}},\ }\href@noop {} {\bibfield  {journal} {\bibinfo
  {journal} {Phys. Rev. Lett.}\ }\textbf {\bibinfo {volume} {102}},\ \bibinfo
  {pages} {040501} (\bibinfo {year} {2009})}\BibitemShut {NoStop}%
\bibitem [{\citenamefont {Monz}\ \emph {et~al.}(2011)\citenamefont {Monz},
  \citenamefont {Schindler}, \citenamefont {Barreiro}, \citenamefont {Chwalla},
  \citenamefont {Nigg}, \citenamefont {Coish}, \citenamefont {Harlander},
  \citenamefont {H\"ansel}, \citenamefont {Hennrich},\ and\ \citenamefont
  {Blatt}}]{Monz2011}%
  \BibitemOpen
  \bibfield  {author} {\bibinfo {author} {\bibfnamefont {T.}~\bibnamefont
  {Monz}}, \bibinfo {author} {\bibfnamefont {P.}~\bibnamefont {Schindler}},
  \bibinfo {author} {\bibfnamefont {J.~T.}\ \bibnamefont {Barreiro}}, \bibinfo
  {author} {\bibfnamefont {M.}~\bibnamefont {Chwalla}}, \bibinfo {author}
  {\bibfnamefont {D.}~\bibnamefont {Nigg}}, \bibinfo {author} {\bibfnamefont
  {W.~A.}\ \bibnamefont {Coish}}, \bibinfo {author} {\bibfnamefont
  {M.}~\bibnamefont {Harlander}}, \bibinfo {author} {\bibfnamefont
  {W.}~\bibnamefont {H\"ansel}}, \bibinfo {author} {\bibfnamefont
  {M.}~\bibnamefont {Hennrich}},\ and\ \bibinfo {author} {\bibfnamefont
  {R.}~\bibnamefont {Blatt}},\ }\href
  {https://doi.org/10.1103/PhysRevLett.106.130506} {\bibfield  {journal}
  {\bibinfo  {journal} {Phys. Rev. Lett.}\ }\textbf {\bibinfo {volume} {106}},\
  \bibinfo {pages} {130506} (\bibinfo {year} {2011})}\BibitemShut {NoStop}%
\bibitem [{\citenamefont {Omran}\ \emph {et~al.}(2019)\citenamefont {Omran},
  \citenamefont {Levine}, \citenamefont {Keesling}, \citenamefont {Semeghini},
  \citenamefont {Wang}, \citenamefont {Ebadi}, \citenamefont {Bernien},
  \citenamefont {Zibrov}, \citenamefont {Pichler}, \citenamefont {Choi} \emph
  {et~al.}}]{omran2019generation}%
  \BibitemOpen
  \bibfield  {author} {\bibinfo {author} {\bibfnamefont {A.}~\bibnamefont
  {Omran}}, \bibinfo {author} {\bibfnamefont {H.}~\bibnamefont {Levine}},
  \bibinfo {author} {\bibfnamefont {A.}~\bibnamefont {Keesling}}, \bibinfo
  {author} {\bibfnamefont {G.}~\bibnamefont {Semeghini}}, \bibinfo {author}
  {\bibfnamefont {T.~T.}\ \bibnamefont {Wang}}, \bibinfo {author}
  {\bibfnamefont {S.}~\bibnamefont {Ebadi}}, \bibinfo {author} {\bibfnamefont
  {H.}~\bibnamefont {Bernien}}, \bibinfo {author} {\bibfnamefont {A.~S.}\
  \bibnamefont {Zibrov}}, \bibinfo {author} {\bibfnamefont {H.}~\bibnamefont
  {Pichler}}, \bibinfo {author} {\bibfnamefont {S.}~\bibnamefont {Choi}}, \emph
  {et~al.},\ }\href@noop {} {\bibfield  {journal} {\bibinfo  {journal}
  {Science}\ }\textbf {\bibinfo {volume} {365}},\ \bibinfo {pages} {570}
  (\bibinfo {year} {2019})}\BibitemShut {NoStop}%
\bibitem [{\citenamefont {Song}\ \emph {et~al.}(2019)\citenamefont {Song},
  \citenamefont {Xu}, \citenamefont {Li}, \citenamefont {Zhang}, \citenamefont
  {Zhang}, \citenamefont {Liu}, \citenamefont {Guo}, \citenamefont {Wang},
  \citenamefont {Ren}, \citenamefont {Hao}, \citenamefont {Feng}, \citenamefont
  {Fan}, \citenamefont {Zheng}, \citenamefont {Wang}, \citenamefont {Wang},\
  and\ \citenamefont {Zhu}}]{Song2019}%
  \BibitemOpen
  \bibfield  {author} {\bibinfo {author} {\bibfnamefont {C.}~\bibnamefont
  {Song}}, \bibinfo {author} {\bibfnamefont {K.}~\bibnamefont {Xu}}, \bibinfo
  {author} {\bibfnamefont {H.}~\bibnamefont {Li}}, \bibinfo {author}
  {\bibfnamefont {Y.-R.}\ \bibnamefont {Zhang}}, \bibinfo {author}
  {\bibfnamefont {X.}~\bibnamefont {Zhang}}, \bibinfo {author} {\bibfnamefont
  {W.}~\bibnamefont {Liu}}, \bibinfo {author} {\bibfnamefont {Q.}~\bibnamefont
  {Guo}}, \bibinfo {author} {\bibfnamefont {Z.}~\bibnamefont {Wang}}, \bibinfo
  {author} {\bibfnamefont {W.}~\bibnamefont {Ren}}, \bibinfo {author}
  {\bibfnamefont {J.}~\bibnamefont {Hao}}, \bibinfo {author} {\bibfnamefont
  {H.}~\bibnamefont {Feng}}, \bibinfo {author} {\bibfnamefont {H.}~\bibnamefont
  {Fan}}, \bibinfo {author} {\bibfnamefont {D.}~\bibnamefont {Zheng}}, \bibinfo
  {author} {\bibfnamefont {D.-W.}\ \bibnamefont {Wang}}, \bibinfo {author}
  {\bibfnamefont {H.}~\bibnamefont {Wang}},\ and\ \bibinfo {author}
  {\bibfnamefont {S.-Y.}\ \bibnamefont {Zhu}},\ }\href
  {https://www.science.org/doi/abs/10.1126/science.aay0600} {\bibfield
  {journal} {\bibinfo  {journal} {Science}\ }\textbf {\bibinfo {volume}
  {365}},\ \bibinfo {pages} {574} (\bibinfo {year} {2019})}\BibitemShut
  {NoStop}%
\bibitem [{\citenamefont {Groenland}\ \emph {et~al.}(2020)\citenamefont
  {Groenland}, \citenamefont {Witteveen}, \citenamefont {Schoutens},\ and\
  \citenamefont {Gerritsma}}]{groenland2020signal}%
  \BibitemOpen
  \bibfield  {author} {\bibinfo {author} {\bibfnamefont {K.}~\bibnamefont
  {Groenland}}, \bibinfo {author} {\bibfnamefont {F.}~\bibnamefont
  {Witteveen}}, \bibinfo {author} {\bibfnamefont {K.}~\bibnamefont
  {Schoutens}},\ and\ \bibinfo {author} {\bibfnamefont {R.}~\bibnamefont
  {Gerritsma}},\ }\href@noop {} {\bibfield  {journal} {\bibinfo  {journal} {New
  Journal of Physics}\ }\textbf {\bibinfo {volume} {22}},\ \bibinfo {pages}
  {063006} (\bibinfo {year} {2020})}\BibitemShut {NoStop}%
\bibitem [{\citenamefont {Toffoli}(1980)}]{Toffoli1980}%
  \BibitemOpen
  \bibfield  {author} {\bibinfo {author} {\bibfnamefont {T.}~\bibnamefont
  {Toffoli}},\ }in\ \href@noop {} {\emph {\bibinfo {booktitle} {Automata,
  Languages and Programming}}},\ \bibinfo {editor} {edited by\ \bibinfo
  {editor} {\bibfnamefont {J.}~\bibnamefont {de~Bakker}}\ and\ \bibinfo
  {editor} {\bibfnamefont {J.}~\bibnamefont {van Leeuwe}}}\ (\bibinfo
  {publisher} {Springer},\ \bibinfo {address} {Berlin, Heidelberg},\ \bibinfo
  {year} {1980})\ pp.\ \bibinfo {pages} {632--644}\BibitemShut {NoStop}%
\bibitem [{\citenamefont {Vedral}\ \emph {et~al.}(1996)\citenamefont {Vedral},
  \citenamefont {Barenco},\ and\ \citenamefont {Ekert}}]{vedral1996quantum}%
  \BibitemOpen
  \bibfield  {author} {\bibinfo {author} {\bibfnamefont {V.}~\bibnamefont
  {Vedral}}, \bibinfo {author} {\bibfnamefont {A.}~\bibnamefont {Barenco}},\
  and\ \bibinfo {author} {\bibfnamefont {A.}~\bibnamefont {Ekert}},\
  }\href@noop {} {\bibfield  {journal} {\bibinfo  {journal} {Phys. Rev. A}\
  }\textbf {\bibinfo {volume} {54}},\ \bibinfo {pages} {147} (\bibinfo {year}
  {1996})}\BibitemShut {NoStop}%
\bibitem [{\citenamefont {Grover}(1996)}]{grover1996fast}%
  \BibitemOpen
  \bibfield  {author} {\bibinfo {author} {\bibfnamefont {L.~K.}\ \bibnamefont
  {Grover}},\ }in\ \href@noop {} {\emph {\bibinfo {booktitle} {Proceedings of
  the twenty-eighth annual ACM symposium on Theory of computing}}}\ (\bibinfo
  {year} {1996})\ pp.\ \bibinfo {pages} {212--219}\BibitemShut {NoStop}%
\bibitem [{\citenamefont {Wang}\ \emph {et~al.}(2001)\citenamefont {Wang},
  \citenamefont {S\o{}rensen},\ and\ \citenamefont {M\o{}lmer}}]{Wang2001}%
  \BibitemOpen
  \bibfield  {author} {\bibinfo {author} {\bibfnamefont {X.}~\bibnamefont
  {Wang}}, \bibinfo {author} {\bibfnamefont {A.}~\bibnamefont {S\o{}rensen}},\
  and\ \bibinfo {author} {\bibfnamefont {K.}~\bibnamefont {M\o{}lmer}},\ }\href
  {https://doi.org/10.1103/PhysRevLett.86.3907} {\bibfield  {journal} {\bibinfo
   {journal} {Phys. Rev. Lett.}\ }\textbf {\bibinfo {volume} {86}},\ \bibinfo
  {pages} {3907} (\bibinfo {year} {2001})}\BibitemShut {NoStop}%
\bibitem [{\citenamefont {Figgatt}\ \emph {et~al.}(2017)\citenamefont
  {Figgatt}, \citenamefont {Maslov}, \citenamefont {Landsman}, \citenamefont
  {Linke}, \citenamefont {Debnath},\ and\ \citenamefont
  {Monroe}}]{figgatt2017complete}%
  \BibitemOpen
  \bibfield  {author} {\bibinfo {author} {\bibfnamefont {C.}~\bibnamefont
  {Figgatt}}, \bibinfo {author} {\bibfnamefont {D.}~\bibnamefont {Maslov}},
  \bibinfo {author} {\bibfnamefont {K.~A.}\ \bibnamefont {Landsman}}, \bibinfo
  {author} {\bibfnamefont {N.~M.}\ \bibnamefont {Linke}}, \bibinfo {author}
  {\bibfnamefont {S.}~\bibnamefont {Debnath}},\ and\ \bibinfo {author}
  {\bibfnamefont {C.}~\bibnamefont {Monroe}},\ }\href@noop {} {\bibfield
  {journal} {\bibinfo  {journal} {Nat. Commun.}\ }\textbf {\bibinfo {volume}
  {8}},\ \bibinfo {pages} {1} (\bibinfo {year} {2017})}\BibitemShut {NoStop}%
\bibitem [{\citenamefont {Paetznick}\ and\ \citenamefont
  {Reichardt}(2013)}]{paetznick2013universal}%
  \BibitemOpen
  \bibfield  {author} {\bibinfo {author} {\bibfnamefont {A.}~\bibnamefont
  {Paetznick}}\ and\ \bibinfo {author} {\bibfnamefont {B.~W.}\ \bibnamefont
  {Reichardt}},\ }\href@noop {} {\bibfield  {journal} {\bibinfo  {journal}
  {Phys. Rev. Lett.}\ }\textbf {\bibinfo {volume} {111}},\ \bibinfo {pages}
  {090505} (\bibinfo {year} {2013})}\BibitemShut {NoStop}%
\bibitem [{\citenamefont {Kitaev}(2003)}]{kitaev2003fault}%
  \BibitemOpen
  \bibfield  {author} {\bibinfo {author} {\bibfnamefont {A.~Y.}\ \bibnamefont
  {Kitaev}},\ }\href@noop {} {\bibfield  {journal} {\bibinfo  {journal} {Annals
  of Physics}\ }\textbf {\bibinfo {volume} {303}},\ \bibinfo {pages} {2}
  (\bibinfo {year} {2003})}\BibitemShut {NoStop}%
\bibitem [{\citenamefont {M{\"u}ller}\ \emph {et~al.}(2011)\citenamefont
  {M{\"u}ller}, \citenamefont {Hammerer}, \citenamefont {Zhou}, \citenamefont
  {Roos},\ and\ \citenamefont {Zoller}}]{muller2011simulating}%
  \BibitemOpen
  \bibfield  {author} {\bibinfo {author} {\bibfnamefont {M.}~\bibnamefont
  {M{\"u}ller}}, \bibinfo {author} {\bibfnamefont {K.}~\bibnamefont
  {Hammerer}}, \bibinfo {author} {\bibfnamefont {Y.}~\bibnamefont {Zhou}},
  \bibinfo {author} {\bibfnamefont {C.~F.}\ \bibnamefont {Roos}},\ and\
  \bibinfo {author} {\bibfnamefont {P.}~\bibnamefont {Zoller}},\ }\href@noop {}
  {\bibfield  {journal} {\bibinfo  {journal} {New J. Phys.}\ }\textbf {\bibinfo
  {volume} {13}},\ \bibinfo {pages} {085007} (\bibinfo {year}
  {2011})}\BibitemShut {NoStop}%
\bibitem [{\citenamefont {O'Malley}\ \emph {et~al.}(2016)\citenamefont
  {O'Malley} \emph {et~al.}}]{o2016scalable}%
  \BibitemOpen
  \bibfield  {author} {\bibinfo {author} {\bibfnamefont {P.~J.~J.}\
  \bibnamefont {O'Malley}} \emph {et~al.},\ }\href@noop {} {\bibfield
  {journal} {\bibinfo  {journal} {Phys. Rev. X}\ }\textbf {\bibinfo {volume}
  {6}},\ \bibinfo {pages} {031007} (\bibinfo {year} {2016})}\BibitemShut
  {NoStop}%
\bibitem [{\citenamefont {Nam}\ \emph {et~al.}(2020)\citenamefont {Nam},
  \citenamefont {Chen}, \citenamefont {Pisenti}, \citenamefont {Wright},
  \citenamefont {Delaney}, \citenamefont {Maslov}, \citenamefont {Brown},
  \citenamefont {Allen}, \citenamefont {Amini}, \citenamefont {Apisdorf} \emph
  {et~al.}}]{nam2020ground}%
  \BibitemOpen
  \bibfield  {author} {\bibinfo {author} {\bibfnamefont {Y.}~\bibnamefont
  {Nam}}, \bibinfo {author} {\bibfnamefont {J.-S.}\ \bibnamefont {Chen}},
  \bibinfo {author} {\bibfnamefont {N.~C.}\ \bibnamefont {Pisenti}}, \bibinfo
  {author} {\bibfnamefont {K.}~\bibnamefont {Wright}}, \bibinfo {author}
  {\bibfnamefont {C.}~\bibnamefont {Delaney}}, \bibinfo {author} {\bibfnamefont
  {D.}~\bibnamefont {Maslov}}, \bibinfo {author} {\bibfnamefont {K.~R.}\
  \bibnamefont {Brown}}, \bibinfo {author} {\bibfnamefont {S.}~\bibnamefont
  {Allen}}, \bibinfo {author} {\bibfnamefont {J.~M.}\ \bibnamefont {Amini}},
  \bibinfo {author} {\bibfnamefont {J.}~\bibnamefont {Apisdorf}}, \emph
  {et~al.},\ }\href@noop {} {\bibfield  {journal} {\bibinfo  {journal} {npj
  Quantum Information}\ }\textbf {\bibinfo {volume} {6}},\ \bibinfo {pages} {1}
  (\bibinfo {year} {2020})}\BibitemShut {NoStop}%
\bibitem [{\citenamefont {Seeley}\ \emph {et~al.}(2012)\citenamefont {Seeley},
  \citenamefont {Richard},\ and\ \citenamefont {Love}}]{seeley2012bravyi}%
  \BibitemOpen
  \bibfield  {author} {\bibinfo {author} {\bibfnamefont {J.~T.}\ \bibnamefont
  {Seeley}}, \bibinfo {author} {\bibfnamefont {M.~J.}\ \bibnamefont
  {Richard}},\ and\ \bibinfo {author} {\bibfnamefont {P.~J.}\ \bibnamefont
  {Love}},\ }\href@noop {} {\bibfield  {journal} {\bibinfo  {journal} {J. Chem.
  Phys.}\ }\textbf {\bibinfo {volume} {137}},\ \bibinfo {pages} {224109}
  (\bibinfo {year} {2012})}\BibitemShut {NoStop}%
\bibitem [{\citenamefont {Banuls}\ \emph {et~al.}(2020)\citenamefont {Banuls},
  \citenamefont {Blatt}, \citenamefont {Catani}, \citenamefont {Celi},
  \citenamefont {Cirac}, \citenamefont {Dalmonte}, \citenamefont {Fallani},
  \citenamefont {Jansen}, \citenamefont {Lewenstein}, \citenamefont
  {Montangero} \emph {et~al.}}]{banuls2020simulating}%
  \BibitemOpen
  \bibfield  {author} {\bibinfo {author} {\bibfnamefont {M.~C.}\ \bibnamefont
  {Banuls}}, \bibinfo {author} {\bibfnamefont {R.}~\bibnamefont {Blatt}},
  \bibinfo {author} {\bibfnamefont {J.}~\bibnamefont {Catani}}, \bibinfo
  {author} {\bibfnamefont {A.}~\bibnamefont {Celi}}, \bibinfo {author}
  {\bibfnamefont {J.~I.}\ \bibnamefont {Cirac}}, \bibinfo {author}
  {\bibfnamefont {M.}~\bibnamefont {Dalmonte}}, \bibinfo {author}
  {\bibfnamefont {L.}~\bibnamefont {Fallani}}, \bibinfo {author} {\bibfnamefont
  {K.}~\bibnamefont {Jansen}}, \bibinfo {author} {\bibfnamefont
  {M.}~\bibnamefont {Lewenstein}}, \bibinfo {author} {\bibfnamefont
  {S.}~\bibnamefont {Montangero}}, \emph {et~al.},\ }\href@noop {} {\bibfield
  {journal} {\bibinfo  {journal} {Eur. Phys. J. D}\ }\textbf {\bibinfo {volume}
  {74}},\ \bibinfo {pages} {1} (\bibinfo {year} {2020})}\BibitemShut {NoStop}%
\bibitem [{\citenamefont {Ciavarella}\ \emph {et~al.}(2021)\citenamefont
  {Ciavarella}, \citenamefont {Klco},\ and\ \citenamefont
  {Savage}}]{ciavarella2021trailhead}%
  \BibitemOpen
  \bibfield  {author} {\bibinfo {author} {\bibfnamefont {A.}~\bibnamefont
  {Ciavarella}}, \bibinfo {author} {\bibfnamefont {N.}~\bibnamefont {Klco}},\
  and\ \bibinfo {author} {\bibfnamefont {M.~J.}\ \bibnamefont {Savage}},\
  }\href@noop {} {\bibfield  {journal} {\bibinfo  {journal} {Phys. Rev. D}\
  }\textbf {\bibinfo {volume} {103}},\ \bibinfo {pages} {094501} (\bibinfo
  {year} {2021})}\BibitemShut {NoStop}%
\bibitem [{\citenamefont {Davoudi}\ \emph {et~al.}(2020)\citenamefont
  {Davoudi}, \citenamefont {Hafezi}, \citenamefont {Monroe}, \citenamefont
  {Pagano}, \citenamefont {Seif},\ and\ \citenamefont
  {Shaw}}]{davoudi2020towards}%
  \BibitemOpen
  \bibfield  {author} {\bibinfo {author} {\bibfnamefont {Z.}~\bibnamefont
  {Davoudi}}, \bibinfo {author} {\bibfnamefont {M.}~\bibnamefont {Hafezi}},
  \bibinfo {author} {\bibfnamefont {C.}~\bibnamefont {Monroe}}, \bibinfo
  {author} {\bibfnamefont {G.}~\bibnamefont {Pagano}}, \bibinfo {author}
  {\bibfnamefont {A.}~\bibnamefont {Seif}},\ and\ \bibinfo {author}
  {\bibfnamefont {A.}~\bibnamefont {Shaw}},\ }\href@noop {} {\bibfield
  {journal} {\bibinfo  {journal} {Physical Review Research}\ }\textbf {\bibinfo
  {volume} {2}},\ \bibinfo {pages} {023015} (\bibinfo {year}
  {2020})}\BibitemShut {NoStop}%
\bibitem [{\citenamefont {Cirac}\ and\ \citenamefont
  {Zoller}(1995)}]{CiracZoller1995}%
  \BibitemOpen
  \bibfield  {author} {\bibinfo {author} {\bibfnamefont {J.~I.}\ \bibnamefont
  {Cirac}}\ and\ \bibinfo {author} {\bibfnamefont {P.}~\bibnamefont {Zoller}},\
  }\href@noop {} {\bibfield  {journal} {\bibinfo  {journal} {Phys. Rev. Lett.}\
  }\textbf {\bibinfo {volume} {74}},\ \bibinfo {pages} {4091} (\bibinfo {year}
  {1995})}\BibitemShut {NoStop}%
\bibitem [{\citenamefont {Goto}\ and\ \citenamefont
  {Ichimura}(2004)}]{goto2004multiqubit}%
  \BibitemOpen
  \bibfield  {author} {\bibinfo {author} {\bibfnamefont {H.}~\bibnamefont
  {Goto}}\ and\ \bibinfo {author} {\bibfnamefont {K.}~\bibnamefont
  {Ichimura}},\ }\href@noop {} {\bibfield  {journal} {\bibinfo  {journal}
  {Phys. Rev. A}\ }\textbf {\bibinfo {volume} {70}},\ \bibinfo {pages} {012305}
  (\bibinfo {year} {2004})}\BibitemShut {NoStop}%
\bibitem [{\citenamefont {Arias~Espinoza}\ \emph {et~al.}(2021)\citenamefont
  {Arias~Espinoza}, \citenamefont {Groenland}, \citenamefont {Mazzanti},
  \citenamefont {Schoutens},\ and\ \citenamefont {Gerritsma}}]{Espinoza2021}%
  \BibitemOpen
  \bibfield  {author} {\bibinfo {author} {\bibfnamefont {J.~D.}\ \bibnamefont
  {Arias~Espinoza}}, \bibinfo {author} {\bibfnamefont {K.}~\bibnamefont
  {Groenland}}, \bibinfo {author} {\bibfnamefont {M.}~\bibnamefont {Mazzanti}},
  \bibinfo {author} {\bibfnamefont {K.}~\bibnamefont {Schoutens}},\ and\
  \bibinfo {author} {\bibfnamefont {R.}~\bibnamefont {Gerritsma}},\ }\href
  {https://doi.org/10.1103/PhysRevA.103.052437} {\bibfield  {journal} {\bibinfo
   {journal} {Phys. Rev. A}\ }\textbf {\bibinfo {volume} {103}},\ \bibinfo
  {pages} {052437} (\bibinfo {year} {2021})}\BibitemShut {NoStop}%
\bibitem [{\citenamefont {Weimer}\ \emph {et~al.}(2010)\citenamefont {Weimer},
  \citenamefont {M{\"u}ller}, \citenamefont {Lesanovsky}, \citenamefont
  {Zoller},\ and\ \citenamefont {B{\"u}chler}}]{weimer2010rydberg}%
  \BibitemOpen
  \bibfield  {author} {\bibinfo {author} {\bibfnamefont {H.}~\bibnamefont
  {Weimer}}, \bibinfo {author} {\bibfnamefont {M.}~\bibnamefont {M{\"u}ller}},
  \bibinfo {author} {\bibfnamefont {I.}~\bibnamefont {Lesanovsky}}, \bibinfo
  {author} {\bibfnamefont {P.}~\bibnamefont {Zoller}},\ and\ \bibinfo {author}
  {\bibfnamefont {H.~P.}\ \bibnamefont {B{\"u}chler}},\ }\href@noop {}
  {\bibfield  {journal} {\bibinfo  {journal} {Nature Physics}\ }\textbf
  {\bibinfo {volume} {6}},\ \bibinfo {pages} {382} (\bibinfo {year}
  {2010})}\BibitemShut {NoStop}%
\bibitem [{\citenamefont {Isenhower}\ \emph {et~al.}(2011)\citenamefont
  {Isenhower}, \citenamefont {Saffman},\ and\ \citenamefont
  {M{\o}lmer}}]{isenhower2011multibit}%
  \BibitemOpen
  \bibfield  {author} {\bibinfo {author} {\bibfnamefont {L.}~\bibnamefont
  {Isenhower}}, \bibinfo {author} {\bibfnamefont {M.}~\bibnamefont {Saffman}},\
  and\ \bibinfo {author} {\bibfnamefont {K.}~\bibnamefont {M{\o}lmer}},\
  }\href@noop {} {\bibfield  {journal} {\bibinfo  {journal} {Quantum Inf.
  Process.}\ }\textbf {\bibinfo {volume} {10}},\ \bibinfo {pages} {755}
  (\bibinfo {year} {2011})}\BibitemShut {NoStop}%
\bibitem [{\citenamefont {Levine}\ \emph {et~al.}(2019)\citenamefont {Levine},
  \citenamefont {Keesling}, \citenamefont {Semeghini}, \citenamefont {Omran},
  \citenamefont {Wang}, \citenamefont {Ebadi}, \citenamefont {Bernien},
  \citenamefont {Greiner}, \citenamefont {Vuleti\ifmmode~\acute{c}\else
  \'{c}\fi{}}, \citenamefont {Pichler},\ and\ \citenamefont
  {Lukin}}]{Levine2019}%
  \BibitemOpen
  \bibfield  {author} {\bibinfo {author} {\bibfnamefont {H.}~\bibnamefont
  {Levine}}, \bibinfo {author} {\bibfnamefont {A.}~\bibnamefont {Keesling}},
  \bibinfo {author} {\bibfnamefont {G.}~\bibnamefont {Semeghini}}, \bibinfo
  {author} {\bibfnamefont {A.}~\bibnamefont {Omran}}, \bibinfo {author}
  {\bibfnamefont {T.~T.}\ \bibnamefont {Wang}}, \bibinfo {author}
  {\bibfnamefont {S.}~\bibnamefont {Ebadi}}, \bibinfo {author} {\bibfnamefont
  {H.}~\bibnamefont {Bernien}}, \bibinfo {author} {\bibfnamefont
  {M.}~\bibnamefont {Greiner}}, \bibinfo {author} {\bibfnamefont
  {V.}~\bibnamefont {Vuleti\ifmmode~\acute{c}\else \'{c}\fi{}}}, \bibinfo
  {author} {\bibfnamefont {H.}~\bibnamefont {Pichler}},\ and\ \bibinfo {author}
  {\bibfnamefont {M.~D.}\ \bibnamefont {Lukin}},\ }\href
  {https://doi.org/10.1103/PhysRevLett.123.170503} {\bibfield  {journal}
  {\bibinfo  {journal} {Phys. Rev. Lett.}\ }\textbf {\bibinfo {volume} {123}},\
  \bibinfo {pages} {170503} (\bibinfo {year} {2019})}\BibitemShut {NoStop}%
\bibitem [{\citenamefont {Khazali}\ and\ \citenamefont
  {M{\o}lmer}(2020)}]{khazali2020fast}%
  \BibitemOpen
  \bibfield  {author} {\bibinfo {author} {\bibfnamefont {M.}~\bibnamefont
  {Khazali}}\ and\ \bibinfo {author} {\bibfnamefont {K.}~\bibnamefont
  {M{\o}lmer}},\ }\href@noop {} {\bibfield  {journal} {\bibinfo  {journal}
  {Phys. Rev. X}\ }\textbf {\bibinfo {volume} {10}},\ \bibinfo {pages} {021054}
  (\bibinfo {year} {2020})}\BibitemShut {NoStop}%
\bibitem [{\citenamefont {Xing}\ \emph {et~al.}(2021)\citenamefont {Xing},
  \citenamefont {Zhao},\ and\ \citenamefont {Tong}}]{xing2021realization}%
  \BibitemOpen
  \bibfield  {author} {\bibinfo {author} {\bibfnamefont {T.~H.}\ \bibnamefont
  {Xing}}, \bibinfo {author} {\bibfnamefont {P.~Z.}\ \bibnamefont {Zhao}},\
  and\ \bibinfo {author} {\bibfnamefont {D.~M.}\ \bibnamefont {Tong}},\
  }\href@noop {} {\bibfield  {journal} {\bibinfo  {journal} {Phys. Rev. A}\
  }\textbf {\bibinfo {volume} {104}},\ \bibinfo {pages} {012618} (\bibinfo
  {year} {2021})}\BibitemShut {NoStop}%
\bibitem [{\citenamefont {Zahedinejad}\ \emph {et~al.}(2015)\citenamefont
  {Zahedinejad}, \citenamefont {Ghosh},\ and\ \citenamefont
  {Sanders}}]{Zahedinejad2015}%
  \BibitemOpen
  \bibfield  {author} {\bibinfo {author} {\bibfnamefont {E.}~\bibnamefont
  {Zahedinejad}}, \bibinfo {author} {\bibfnamefont {J.}~\bibnamefont {Ghosh}},\
  and\ \bibinfo {author} {\bibfnamefont {B.~C.}\ \bibnamefont {Sanders}},\
  }\href {https://doi.org/10.1103/PhysRevLett.114.200502} {\bibfield  {journal}
  {\bibinfo  {journal} {Phys. Rev. Lett.}\ }\textbf {\bibinfo {volume} {114}},\
  \bibinfo {pages} {200502} (\bibinfo {year} {2015})}\BibitemShut {NoStop}%
\bibitem [{\citenamefont {Wineland}\ \emph {et~al.}(1998)\citenamefont
  {Wineland}, \citenamefont {Monroe}, \citenamefont {Itano}, \citenamefont
  {Leibfried}, \citenamefont {King},\ and\ \citenamefont
  {Meekhof}}]{wineland1998experimental}%
  \BibitemOpen
  \bibfield  {author} {\bibinfo {author} {\bibfnamefont {D.~J.}\ \bibnamefont
  {Wineland}}, \bibinfo {author} {\bibfnamefont {C.}~\bibnamefont {Monroe}},
  \bibinfo {author} {\bibfnamefont {W.~M.}\ \bibnamefont {Itano}}, \bibinfo
  {author} {\bibfnamefont {D.}~\bibnamefont {Leibfried}}, \bibinfo {author}
  {\bibfnamefont {B.~E.}\ \bibnamefont {King}},\ and\ \bibinfo {author}
  {\bibfnamefont {D.~M.}\ \bibnamefont {Meekhof}},\ }\href@noop {} {\bibfield
  {journal} {\bibinfo  {journal} {J. Res. Nat. Inst. Stand. Technol.}\ }\textbf
  {\bibinfo {volume} {103}},\ \bibinfo {pages} {259} (\bibinfo {year}
  {1998})}\BibitemShut {NoStop}%
\bibitem [{\citenamefont {Brown}\ \emph {et~al.}(2016)\citenamefont {Brown},
  \citenamefont {Kim},\ and\ \citenamefont {Monroe}}]{Brown2016}%
  \BibitemOpen
  \bibfield  {author} {\bibinfo {author} {\bibfnamefont {K.~R.}\ \bibnamefont
  {Brown}}, \bibinfo {author} {\bibfnamefont {J.}~\bibnamefont {Kim}},\ and\
  \bibinfo {author} {\bibfnamefont {C.}~\bibnamefont {Monroe}},\ }\href
  {https://doi.org/10.1038/npjqi.2016.34} {\bibfield  {journal} {\bibinfo
  {journal} {npj Quantum Inf.}\ }\textbf {\bibinfo {volume} {2}},\ \bibinfo
  {pages} {16034} (\bibinfo {year} {2016})}\BibitemShut {NoStop}%
\bibitem [{\citenamefont {Lubinski}\ \emph {et~al.}(2021)\citenamefont
  {Lubinski}, \citenamefont {Johri}, \citenamefont {Varosy}, \citenamefont
  {Coleman}, \citenamefont {Zhao}, \citenamefont {Necaise}, \citenamefont
  {Baldwin}, \citenamefont {Mayer},\ and\ \citenamefont {Proctor}}]{QEDC}%
  \BibitemOpen
  \bibfield  {author} {\bibinfo {author} {\bibfnamefont {T.}~\bibnamefont
  {Lubinski}}, \bibinfo {author} {\bibfnamefont {S.}~\bibnamefont {Johri}},
  \bibinfo {author} {\bibfnamefont {P.}~\bibnamefont {Varosy}}, \bibinfo
  {author} {\bibfnamefont {J.}~\bibnamefont {Coleman}}, \bibinfo {author}
  {\bibfnamefont {L.}~\bibnamefont {Zhao}}, \bibinfo {author} {\bibfnamefont
  {J.}~\bibnamefont {Necaise}}, \bibinfo {author} {\bibfnamefont {C.~H.}\
  \bibnamefont {Baldwin}}, \bibinfo {author} {\bibfnamefont {K.}~\bibnamefont
  {Mayer}},\ and\ \bibinfo {author} {\bibfnamefont {T.}~\bibnamefont
  {Proctor}},\ }\href@noop {} {\bibfield  {journal} {\bibinfo  {journal}
  {arXiv:2110.03137}\ } (\bibinfo {year} {2021})}\BibitemShut {NoStop}%
\bibitem [{\citenamefont {M{\o}lmer}\ and\ \citenamefont
  {S{\o}rensen}(1999)}]{Molmer1999}%
  \BibitemOpen
  \bibfield  {author} {\bibinfo {author} {\bibfnamefont {K.}~\bibnamefont
  {M{\o}lmer}}\ and\ \bibinfo {author} {\bibfnamefont {A.}~\bibnamefont
  {S{\o}rensen}},\ }\href@noop {} {\bibfield  {journal} {\bibinfo  {journal}
  {Phys. Rev. Lett.}\ }\textbf {\bibinfo {volume} {82}},\ \bibinfo {pages}
  {1835} (\bibinfo {year} {1999})}\BibitemShut {NoStop}%
\bibitem [{\citenamefont {S\o{}rensen}\ and\ \citenamefont
  {M\o{}lmer}(1999)}]{Sorensen1999}%
  \BibitemOpen
  \bibfield  {author} {\bibinfo {author} {\bibfnamefont {A.}~\bibnamefont
  {S\o{}rensen}}\ and\ \bibinfo {author} {\bibfnamefont {K.}~\bibnamefont
  {M\o{}lmer}},\ }\href@noop {} {\bibfield  {journal} {\bibinfo  {journal}
  {Phys. Rev. Lett.}\ }\textbf {\bibinfo {volume} {82}},\ \bibinfo {pages}
  {1971} (\bibinfo {year} {1999})}\BibitemShut {NoStop}%
\bibitem [{\citenamefont {Solano}\ \emph {et~al.}(1999)\citenamefont {Solano},
  \citenamefont {de~Matos~Filho},\ and\ \citenamefont {Zagury}}]{Solano1999}%
  \BibitemOpen
  \bibfield  {author} {\bibinfo {author} {\bibfnamefont {E.}~\bibnamefont
  {Solano}}, \bibinfo {author} {\bibfnamefont {R.~L.}\ \bibnamefont
  {de~Matos~Filho}},\ and\ \bibinfo {author} {\bibfnamefont {N.}~\bibnamefont
  {Zagury}},\ }\href {https://doi.org/10.1103/PhysRevA.59.R2539} {\bibfield
  {journal} {\bibinfo  {journal} {Phys. Rev. A}\ }\textbf {\bibinfo {volume}
  {59}},\ \bibinfo {pages} {R2539} (\bibinfo {year} {1999})}\BibitemShut
  {NoStop}%
\bibitem [{\citenamefont {Milburn}\ \emph {et~al.}(2000)\citenamefont
  {Milburn}, \citenamefont {Schneider},\ and\ \citenamefont
  {James}}]{Milburn2000}%
  \BibitemOpen
  \bibfield  {author} {\bibinfo {author} {\bibfnamefont {G.}~\bibnamefont
  {Milburn}}, \bibinfo {author} {\bibfnamefont {S.}~\bibnamefont {Schneider}},\
  and\ \bibinfo {author} {\bibfnamefont {D.}~\bibnamefont {James}},\
  }\href@noop {} {\bibfield  {journal} {\bibinfo  {journal} {Fortschritte der
  Phys.}\ }\textbf {\bibinfo {volume} {48}},\ \bibinfo {pages} {801} (\bibinfo
  {year} {2000})}\BibitemShut {NoStop}%
\bibitem [{\citenamefont {Drechsler}\ \emph {et~al.}(2020)\citenamefont
  {Drechsler}, \citenamefont {Far{\'\i}as}, \citenamefont {Freitas},
  \citenamefont {Schmiegelow},\ and\ \citenamefont {Paz}}]{drechsler2020state}%
  \BibitemOpen
  \bibfield  {author} {\bibinfo {author} {\bibfnamefont {M.}~\bibnamefont
  {Drechsler}}, \bibinfo {author} {\bibfnamefont {M.~B.}\ \bibnamefont
  {Far{\'\i}as}}, \bibinfo {author} {\bibfnamefont {N.}~\bibnamefont
  {Freitas}}, \bibinfo {author} {\bibfnamefont {C.~T.}\ \bibnamefont
  {Schmiegelow}},\ and\ \bibinfo {author} {\bibfnamefont {J.~P.}\ \bibnamefont
  {Paz}},\ }\href@noop {} {\bibfield  {journal} {\bibinfo  {journal} {Physical
  Review A}\ }\textbf {\bibinfo {volume} {101}},\ \bibinfo {pages} {052331}
  (\bibinfo {year} {2020})}\BibitemShut {NoStop}%
\bibitem [{\citenamefont {Sutherland}\ \emph {et~al.}(2021)\citenamefont
  {Sutherland}, \citenamefont {Burd}, \citenamefont {Slichter}, \citenamefont
  {Libby},\ and\ \citenamefont {Leibfried}}]{sutherland2021motional}%
  \BibitemOpen
  \bibfield  {author} {\bibinfo {author} {\bibfnamefont {R.~T.}\ \bibnamefont
  {Sutherland}}, \bibinfo {author} {\bibfnamefont {S.}~\bibnamefont {Burd}},
  \bibinfo {author} {\bibfnamefont {D.}~\bibnamefont {Slichter}}, \bibinfo
  {author} {\bibfnamefont {S.}~\bibnamefont {Libby}},\ and\ \bibinfo {author}
  {\bibfnamefont {D.}~\bibnamefont {Leibfried}},\ }\href@noop {} {\bibfield
  {journal} {\bibinfo  {journal} {Physical Review Letters}\ }\textbf {\bibinfo
  {volume} {127}},\ \bibinfo {pages} {083201} (\bibinfo {year}
  {2021})}\BibitemShut {NoStop}%
\bibitem [{\citenamefont {Bohnet}\ \emph {et~al.}(2016)\citenamefont {Bohnet},
  \citenamefont {Sawyer}, \citenamefont {Britton}, \citenamefont {Wall},
  \citenamefont {Rey}, \citenamefont {Foss-Feig},\ and\ \citenamefont
  {Bollinger}}]{bohnet2016quantum}%
  \BibitemOpen
  \bibfield  {author} {\bibinfo {author} {\bibfnamefont {J.~G.}\ \bibnamefont
  {Bohnet}}, \bibinfo {author} {\bibfnamefont {B.~C.}\ \bibnamefont {Sawyer}},
  \bibinfo {author} {\bibfnamefont {J.~W.}\ \bibnamefont {Britton}}, \bibinfo
  {author} {\bibfnamefont {M.~L.}\ \bibnamefont {Wall}}, \bibinfo {author}
  {\bibfnamefont {A.~M.}\ \bibnamefont {Rey}}, \bibinfo {author} {\bibfnamefont
  {M.}~\bibnamefont {Foss-Feig}},\ and\ \bibinfo {author} {\bibfnamefont
  {J.~J.}\ \bibnamefont {Bollinger}},\ }\href@noop {} {\bibfield  {journal}
  {\bibinfo  {journal} {Science}\ }\textbf {\bibinfo {volume} {352}},\ \bibinfo
  {pages} {1297} (\bibinfo {year} {2016})}\BibitemShut {NoStop}%
\bibitem [{\citenamefont {Feng}\ \emph {et~al.}(2021)\citenamefont {Feng},
  \citenamefont {Ilo-Okeke}, \citenamefont {Pyrkov}, \citenamefont
  {Askitopoulos},\ and\ \citenamefont {Byrnes}}]{feng2021sensitive}%
  \BibitemOpen
  \bibfield  {author} {\bibinfo {author} {\bibfnamefont {J.}~\bibnamefont
  {Feng}}, \bibinfo {author} {\bibfnamefont {E.~O.}\ \bibnamefont {Ilo-Okeke}},
  \bibinfo {author} {\bibfnamefont {A.~N.}\ \bibnamefont {Pyrkov}}, \bibinfo
  {author} {\bibfnamefont {A.}~\bibnamefont {Askitopoulos}},\ and\ \bibinfo
  {author} {\bibfnamefont {T.}~\bibnamefont {Byrnes}},\ }\href@noop {}
  {\bibfield  {journal} {\bibinfo  {journal} {Physical Review A}\ }\textbf
  {\bibinfo {volume} {104}},\ \bibinfo {pages} {013318} (\bibinfo {year}
  {2021})}\BibitemShut {NoStop}%
\bibitem [{\citenamefont {Ge}\ \emph {et~al.}(2019{\natexlab{a}})\citenamefont
  {Ge}, \citenamefont {Sawyer}, \citenamefont {Britton}, \citenamefont
  {Jacobs}, \citenamefont {Bollinger},\ and\ \citenamefont
  {Foss-Feig}}]{ge2019trapped}%
  \BibitemOpen
  \bibfield  {author} {\bibinfo {author} {\bibfnamefont {W.}~\bibnamefont
  {Ge}}, \bibinfo {author} {\bibfnamefont {B.~C.}\ \bibnamefont {Sawyer}},
  \bibinfo {author} {\bibfnamefont {J.~W.}\ \bibnamefont {Britton}}, \bibinfo
  {author} {\bibfnamefont {K.}~\bibnamefont {Jacobs}}, \bibinfo {author}
  {\bibfnamefont {J.~J.}\ \bibnamefont {Bollinger}},\ and\ \bibinfo {author}
  {\bibfnamefont {M.}~\bibnamefont {Foss-Feig}},\ }\href@noop {} {\bibfield
  {journal} {\bibinfo  {journal} {Phys. Rev. Lett.}\ }\textbf {\bibinfo
  {volume} {122}},\ \bibinfo {pages} {030501} (\bibinfo {year}
  {2019}{\natexlab{a}})}\BibitemShut {NoStop}%
\bibitem [{\citenamefont {Ge}\ \emph {et~al.}(2019{\natexlab{b}})\citenamefont
  {Ge}, \citenamefont {Sawyer}, \citenamefont {Britton}, \citenamefont
  {Jacobs}, \citenamefont {Foss-Feig},\ and\ \citenamefont
  {Bollinger}}]{ge2019stroboscopic}%
  \BibitemOpen
  \bibfield  {author} {\bibinfo {author} {\bibfnamefont {W.}~\bibnamefont
  {Ge}}, \bibinfo {author} {\bibfnamefont {B.~C.}\ \bibnamefont {Sawyer}},
  \bibinfo {author} {\bibfnamefont {J.~W.}\ \bibnamefont {Britton}}, \bibinfo
  {author} {\bibfnamefont {K.}~\bibnamefont {Jacobs}}, \bibinfo {author}
  {\bibfnamefont {M.}~\bibnamefont {Foss-Feig}},\ and\ \bibinfo {author}
  {\bibfnamefont {J.~J.}\ \bibnamefont {Bollinger}},\ }\href@noop {} {\bibfield
   {journal} {\bibinfo  {journal} {Phys. Rev. A}\ }\textbf {\bibinfo {volume}
  {100}},\ \bibinfo {pages} {043417} (\bibinfo {year}
  {2019}{\natexlab{b}})}\BibitemShut {NoStop}%
\bibitem [{\citenamefont {Burd}\ \emph {et~al.}(2021)\citenamefont {Burd},
  \citenamefont {Srinivas}, \citenamefont {Knaack}, \citenamefont {Ge},
  \citenamefont {Wilson}, \citenamefont {Wineland}, \citenamefont {Leibfried},
  \citenamefont {Bollinger}, \citenamefont {Allcock},\ and\ \citenamefont
  {Slichter}}]{burd2021quantum}%
  \BibitemOpen
  \bibfield  {author} {\bibinfo {author} {\bibfnamefont {S.~C.}\ \bibnamefont
  {Burd}}, \bibinfo {author} {\bibfnamefont {R.}~\bibnamefont {Srinivas}},
  \bibinfo {author} {\bibfnamefont {H.~M.}\ \bibnamefont {Knaack}}, \bibinfo
  {author} {\bibfnamefont {W.}~\bibnamefont {Ge}}, \bibinfo {author}
  {\bibfnamefont {A.~C.}\ \bibnamefont {Wilson}}, \bibinfo {author}
  {\bibfnamefont {D.~J.}\ \bibnamefont {Wineland}}, \bibinfo {author}
  {\bibfnamefont {D.}~\bibnamefont {Leibfried}}, \bibinfo {author}
  {\bibfnamefont {J.~J.}\ \bibnamefont {Bollinger}}, \bibinfo {author}
  {\bibfnamefont {D.}~\bibnamefont {Allcock}},\ and\ \bibinfo {author}
  {\bibfnamefont {D.}~\bibnamefont {Slichter}},\ }\href@noop {} {\bibfield
  {journal} {\bibinfo  {journal} {Nature Physics}\ }\textbf {\bibinfo {volume}
  {17}},\ \bibinfo {pages} {898} (\bibinfo {year} {2021})}\BibitemShut
  {NoStop}%
\bibitem [{\citenamefont {Leibfried}\ \emph {et~al.}(2003)\citenamefont
  {Leibfried}, \citenamefont {Blatt}, \citenamefont {Monroe},\ and\
  \citenamefont {Wineland}}]{Leibfried2003}%
  \BibitemOpen
  \bibfield  {author} {\bibinfo {author} {\bibfnamefont {D.}~\bibnamefont
  {Leibfried}}, \bibinfo {author} {\bibfnamefont {R.}~\bibnamefont {Blatt}},
  \bibinfo {author} {\bibfnamefont {C.}~\bibnamefont {Monroe}},\ and\ \bibinfo
  {author} {\bibfnamefont {D.}~\bibnamefont {Wineland}},\ }\href@noop {}
  {\bibfield  {journal} {\bibinfo  {journal} {Rev. Mod. Phys.}\ }\textbf
  {\bibinfo {volume} {75}},\ \bibinfo {pages} {281} (\bibinfo {year}
  {2003})}\BibitemShut {NoStop}%
\bibitem [{\citenamefont {S{\o}rensen}\ and\ \citenamefont
  {M{\o}lmer}(2000)}]{sorensen2000entanglement}%
  \BibitemOpen
  \bibfield  {author} {\bibinfo {author} {\bibfnamefont {A.}~\bibnamefont
  {S{\o}rensen}}\ and\ \bibinfo {author} {\bibfnamefont {K.}~\bibnamefont
  {M{\o}lmer}},\ }\href@noop {} {\bibfield  {journal} {\bibinfo  {journal}
  {Phys. Rev. A}\ }\textbf {\bibinfo {volume} {62}},\ \bibinfo {pages} {022311}
  (\bibinfo {year} {2000})}\BibitemShut {NoStop}%
\bibitem [{\citenamefont {Zhu}\ \emph {et~al.}(2006)\citenamefont {Zhu},
  \citenamefont {Monroe},\ and\ \citenamefont {Duan}}]{Zhu2006}%
  \BibitemOpen
  \bibfield  {author} {\bibinfo {author} {\bibfnamefont {S.-L.}\ \bibnamefont
  {Zhu}}, \bibinfo {author} {\bibfnamefont {C.}~\bibnamefont {Monroe}},\ and\
  \bibinfo {author} {\bibfnamefont {L.-M.}\ \bibnamefont {Duan}},\ }\href
  {https://doi.org/10.1103/PhysRevLett.97.050505} {\bibfield  {journal}
  {\bibinfo  {journal} {Phys. Rev. Lett.}\ }\textbf {\bibinfo {volume} {97}},\
  \bibinfo {pages} {050505} (\bibinfo {year} {2006})}\BibitemShut {NoStop}%
\bibitem [{\citenamefont {Debnath}\ \emph {et~al.}(2016)\citenamefont
  {Debnath}, \citenamefont {Linke}, \citenamefont {Figgatt}, \citenamefont
  {Landsman}, \citenamefont {Wright},\ and\ \citenamefont
  {Monroe}}]{Debnath2016}%
  \BibitemOpen
  \bibfield  {author} {\bibinfo {author} {\bibfnamefont {S.}~\bibnamefont
  {Debnath}}, \bibinfo {author} {\bibfnamefont {N.~M.}\ \bibnamefont {Linke}},
  \bibinfo {author} {\bibfnamefont {C.}~\bibnamefont {Figgatt}}, \bibinfo
  {author} {\bibfnamefont {K.~A.}\ \bibnamefont {Landsman}}, \bibinfo {author}
  {\bibfnamefont {K.}~\bibnamefont {Wright}},\ and\ \bibinfo {author}
  {\bibfnamefont {C.}~\bibnamefont {Monroe}},\ }\href
  {https://doi.org/10.1038/nature18648} {\bibfield  {journal} {\bibinfo
  {journal} {Nature}\ }\textbf {\bibinfo {volume} {536}},\ \bibinfo {pages}
  {63} (\bibinfo {year} {2016})}\BibitemShut {NoStop}%
\bibitem [{\citenamefont {Lu}\ \emph {et~al.}(2019)\citenamefont {Lu},
  \citenamefont {Zhang}, \citenamefont {Zhang}, \citenamefont {Chen},
  \citenamefont {Shen}, \citenamefont {Zhang}, \citenamefont {Zhang},\ and\
  \citenamefont {Kim}}]{lu2019global}%
  \BibitemOpen
  \bibfield  {author} {\bibinfo {author} {\bibfnamefont {Y.}~\bibnamefont
  {Lu}}, \bibinfo {author} {\bibfnamefont {S.}~\bibnamefont {Zhang}}, \bibinfo
  {author} {\bibfnamefont {K.}~\bibnamefont {Zhang}}, \bibinfo {author}
  {\bibfnamefont {W.}~\bibnamefont {Chen}}, \bibinfo {author} {\bibfnamefont
  {Y.}~\bibnamefont {Shen}}, \bibinfo {author} {\bibfnamefont {J.}~\bibnamefont
  {Zhang}}, \bibinfo {author} {\bibfnamefont {J.-N.}\ \bibnamefont {Zhang}},\
  and\ \bibinfo {author} {\bibfnamefont {K.}~\bibnamefont {Kim}},\ }\href@noop
  {} {\bibfield  {journal} {\bibinfo  {journal} {Nature}\ }\textbf {\bibinfo
  {volume} {572}},\ \bibinfo {pages} {363} (\bibinfo {year}
  {2019})}\BibitemShut {NoStop}%
\bibitem [{\citenamefont {Figgatt}\ \emph {et~al.}(2019)\citenamefont
  {Figgatt}, \citenamefont {Ostrander}, \citenamefont {Linke}, \citenamefont
  {Landsman}, \citenamefont {Zhu}, \citenamefont {Maslov},\ and\ \citenamefont
  {Monroe}}]{figgatt2019parallel}%
  \BibitemOpen
  \bibfield  {author} {\bibinfo {author} {\bibfnamefont {C.}~\bibnamefont
  {Figgatt}}, \bibinfo {author} {\bibfnamefont {A.}~\bibnamefont {Ostrander}},
  \bibinfo {author} {\bibfnamefont {N.~M.}\ \bibnamefont {Linke}}, \bibinfo
  {author} {\bibfnamefont {K.~A.}\ \bibnamefont {Landsman}}, \bibinfo {author}
  {\bibfnamefont {D.}~\bibnamefont {Zhu}}, \bibinfo {author} {\bibfnamefont
  {D.}~\bibnamefont {Maslov}},\ and\ \bibinfo {author} {\bibfnamefont
  {C.}~\bibnamefont {Monroe}},\ }\href@noop {} {\bibfield  {journal} {\bibinfo
  {journal} {Nature}\ }\textbf {\bibinfo {volume} {572}},\ \bibinfo {pages}
  {368} (\bibinfo {year} {2019})}\BibitemShut {NoStop}%
\bibitem [{\citenamefont {Marvian}(2022)}]{marvian2022restrictions}%
  \BibitemOpen
  \bibfield  {author} {\bibinfo {author} {\bibfnamefont {I.}~\bibnamefont
  {Marvian}},\ }\href@noop {} {\bibfield  {journal} {\bibinfo  {journal}
  {Nature Physics}\ ,\ \bibinfo {pages} {1}} (\bibinfo {year}
  {2022})}\BibitemShut {NoStop}%
\bibitem [{\citenamefont {Roggero}\ \emph {et~al.}(2020)\citenamefont
  {Roggero}, \citenamefont {Li}, \citenamefont {Carlson}, \citenamefont
  {Gupta},\ and\ \citenamefont {Perdue}}]{roggero2020quantum}%
  \BibitemOpen
  \bibfield  {author} {\bibinfo {author} {\bibfnamefont {A.}~\bibnamefont
  {Roggero}}, \bibinfo {author} {\bibfnamefont {A.~C.}\ \bibnamefont {Li}},
  \bibinfo {author} {\bibfnamefont {J.}~\bibnamefont {Carlson}}, \bibinfo
  {author} {\bibfnamefont {R.}~\bibnamefont {Gupta}},\ and\ \bibinfo {author}
  {\bibfnamefont {G.~N.}\ \bibnamefont {Perdue}},\ }\href@noop {} {\bibfield
  {journal} {\bibinfo  {journal} {Physical Review D}\ }\textbf {\bibinfo
  {volume} {101}},\ \bibinfo {pages} {074038} (\bibinfo {year}
  {2020})}\BibitemShut {NoStop}%
\bibitem [{\citenamefont {Pachos}\ and\ \citenamefont
  {Plenio}(2004)}]{pachos2004three}%
  \BibitemOpen
  \bibfield  {author} {\bibinfo {author} {\bibfnamefont {J.~K.}\ \bibnamefont
  {Pachos}}\ and\ \bibinfo {author} {\bibfnamefont {M.~B.}\ \bibnamefont
  {Plenio}},\ }\href@noop {} {\bibfield  {journal} {\bibinfo  {journal}
  {Physical review letters}\ }\textbf {\bibinfo {volume} {93}},\ \bibinfo
  {pages} {056402} (\bibinfo {year} {2004})}\BibitemShut {NoStop}%
\bibitem [{Note1()}]{Note1}%
  \BibitemOpen
  \bibinfo {note} {We calculate the overlap as the entanglement fidelity of the
  two unitary processes. c.f.~\cite {horodecki1999general}.}\BibitemShut
  {Stop}%
\bibitem [{\citenamefont {Zhou}\ \emph {et~al.}(2020)\citenamefont {Zhou},
  \citenamefont {Wang}, \citenamefont {Choi}, \citenamefont {Pichler},\ and\
  \citenamefont {Lukin}}]{zhou2020quantum}%
  \BibitemOpen
  \bibfield  {author} {\bibinfo {author} {\bibfnamefont {L.}~\bibnamefont
  {Zhou}}, \bibinfo {author} {\bibfnamefont {S.-T.}\ \bibnamefont {Wang}},
  \bibinfo {author} {\bibfnamefont {S.}~\bibnamefont {Choi}}, \bibinfo {author}
  {\bibfnamefont {H.}~\bibnamefont {Pichler}},\ and\ \bibinfo {author}
  {\bibfnamefont {M.~D.}\ \bibnamefont {Lukin}},\ }\href@noop {} {\bibfield
  {journal} {\bibinfo  {journal} {Physical Review X}\ }\textbf {\bibinfo
  {volume} {10}},\ \bibinfo {pages} {021067} (\bibinfo {year}
  {2020})}\BibitemShut {NoStop}%
\bibitem [{\citenamefont {Cerezo}\ \emph {et~al.}(2021)\citenamefont {Cerezo},
  \citenamefont {Arrasmith}, \citenamefont {Babbush}, \citenamefont {Benjamin},
  \citenamefont {Endo}, \citenamefont {Fujii}, \citenamefont {McClean},
  \citenamefont {Mitarai}, \citenamefont {Yuan}, \citenamefont {Cincio} \emph
  {et~al.}}]{cerezo2021variational}%
  \BibitemOpen
  \bibfield  {author} {\bibinfo {author} {\bibfnamefont {M.}~\bibnamefont
  {Cerezo}}, \bibinfo {author} {\bibfnamefont {A.}~\bibnamefont {Arrasmith}},
  \bibinfo {author} {\bibfnamefont {R.}~\bibnamefont {Babbush}}, \bibinfo
  {author} {\bibfnamefont {S.~C.}\ \bibnamefont {Benjamin}}, \bibinfo {author}
  {\bibfnamefont {S.}~\bibnamefont {Endo}}, \bibinfo {author} {\bibfnamefont
  {K.}~\bibnamefont {Fujii}}, \bibinfo {author} {\bibfnamefont {J.~R.}\
  \bibnamefont {McClean}}, \bibinfo {author} {\bibfnamefont {K.}~\bibnamefont
  {Mitarai}}, \bibinfo {author} {\bibfnamefont {X.}~\bibnamefont {Yuan}},
  \bibinfo {author} {\bibfnamefont {L.}~\bibnamefont {Cincio}}, \emph
  {et~al.},\ }\href@noop {} {\bibfield  {journal} {\bibinfo  {journal} {Nature
  Reviews Physics}\ }\textbf {\bibinfo {volume} {3}},\ \bibinfo {pages} {625}
  (\bibinfo {year} {2021})}\BibitemShut {NoStop}%
\bibitem [{\citenamefont {Kandala}\ \emph {et~al.}(2017)\citenamefont
  {Kandala}, \citenamefont {Mezzacapo}, \citenamefont {Temme}, \citenamefont
  {Takita}, \citenamefont {Brink}, \citenamefont {Chow},\ and\ \citenamefont
  {Gambetta}}]{kandala2017hardware}%
  \BibitemOpen
  \bibfield  {author} {\bibinfo {author} {\bibfnamefont {A.}~\bibnamefont
  {Kandala}}, \bibinfo {author} {\bibfnamefont {A.}~\bibnamefont {Mezzacapo}},
  \bibinfo {author} {\bibfnamefont {K.}~\bibnamefont {Temme}}, \bibinfo
  {author} {\bibfnamefont {M.}~\bibnamefont {Takita}}, \bibinfo {author}
  {\bibfnamefont {M.}~\bibnamefont {Brink}}, \bibinfo {author} {\bibfnamefont
  {J.~M.}\ \bibnamefont {Chow}},\ and\ \bibinfo {author} {\bibfnamefont
  {J.~M.}\ \bibnamefont {Gambetta}},\ }\href@noop {} {\bibfield  {journal}
  {\bibinfo  {journal} {Nature}\ }\textbf {\bibinfo {volume} {549}},\ \bibinfo
  {pages} {242} (\bibinfo {year} {2017})}\BibitemShut {NoStop}%
\bibitem [{\citenamefont {Monroe}\ \emph {et~al.}(2021)\citenamefont {Monroe},
  \citenamefont {Campbell}, \citenamefont {Duan}, \citenamefont {Gong},
  \citenamefont {Gorshkov}, \citenamefont {Hess}, \citenamefont {Islam},
  \citenamefont {Kim}, \citenamefont {Linke}, \citenamefont {Pagano},
  \citenamefont {Richerme}, \citenamefont {Senko},\ and\ \citenamefont
  {Yao}}]{Monroe2021}%
  \BibitemOpen
  \bibfield  {author} {\bibinfo {author} {\bibfnamefont {C.}~\bibnamefont
  {Monroe}}, \bibinfo {author} {\bibfnamefont {W.~C.}\ \bibnamefont
  {Campbell}}, \bibinfo {author} {\bibfnamefont {L.-M.}\ \bibnamefont {Duan}},
  \bibinfo {author} {\bibfnamefont {Z.-X.}\ \bibnamefont {Gong}}, \bibinfo
  {author} {\bibfnamefont {A.~V.}\ \bibnamefont {Gorshkov}}, \bibinfo {author}
  {\bibfnamefont {P.~W.}\ \bibnamefont {Hess}}, \bibinfo {author}
  {\bibfnamefont {R.}~\bibnamefont {Islam}}, \bibinfo {author} {\bibfnamefont
  {K.}~\bibnamefont {Kim}}, \bibinfo {author} {\bibfnamefont {N.~M.}\
  \bibnamefont {Linke}}, \bibinfo {author} {\bibfnamefont {G.}~\bibnamefont
  {Pagano}}, \bibinfo {author} {\bibfnamefont {P.}~\bibnamefont {Richerme}},
  \bibinfo {author} {\bibfnamefont {C.}~\bibnamefont {Senko}},\ and\ \bibinfo
  {author} {\bibfnamefont {N.~Y.}\ \bibnamefont {Yao}},\ }\href
  {https://doi.org/10.1103/RevModPhys.93.025001} {\bibfield  {journal}
  {\bibinfo  {journal} {Rev. Mod. Phys.}\ }\textbf {\bibinfo {volume} {93}},\
  \bibinfo {pages} {025001} (\bibinfo {year} {2021})}\BibitemShut {NoStop}%
\bibitem [{\citenamefont {Gilmore}\ \emph {et~al.}(2021)\citenamefont
  {Gilmore}, \citenamefont {Affolter}, \citenamefont {Lewis-Swan},
  \citenamefont {Barberena}, \citenamefont {Jordan}, \citenamefont {Rey},\ and\
  \citenamefont {Bollinger}}]{gilmore2021quantum}%
  \BibitemOpen
  \bibfield  {author} {\bibinfo {author} {\bibfnamefont {K.~A.}\ \bibnamefont
  {Gilmore}}, \bibinfo {author} {\bibfnamefont {M.}~\bibnamefont {Affolter}},
  \bibinfo {author} {\bibfnamefont {R.~J.}\ \bibnamefont {Lewis-Swan}},
  \bibinfo {author} {\bibfnamefont {D.}~\bibnamefont {Barberena}}, \bibinfo
  {author} {\bibfnamefont {E.}~\bibnamefont {Jordan}}, \bibinfo {author}
  {\bibfnamefont {A.~M.}\ \bibnamefont {Rey}},\ and\ \bibinfo {author}
  {\bibfnamefont {J.~J.}\ \bibnamefont {Bollinger}},\ }\href@noop {} {\bibfield
   {journal} {\bibinfo  {journal} {Science}\ }\textbf {\bibinfo {volume}
  {373}},\ \bibinfo {pages} {673} (\bibinfo {year} {2021})}\BibitemShut
  {NoStop}%
\bibitem [{\citenamefont {Haljan}\ \emph {et~al.}(2005)\citenamefont {Haljan},
  \citenamefont {Brickman}, \citenamefont {Deslauriers}, \citenamefont {Lee},\
  and\ \citenamefont {Monroe}}]{haljan2005spin}%
  \BibitemOpen
  \bibfield  {author} {\bibinfo {author} {\bibfnamefont {P.~C.}\ \bibnamefont
  {Haljan}}, \bibinfo {author} {\bibfnamefont {K.-A.}\ \bibnamefont
  {Brickman}}, \bibinfo {author} {\bibfnamefont {L.}~\bibnamefont
  {Deslauriers}}, \bibinfo {author} {\bibfnamefont {P.~J.}\ \bibnamefont
  {Lee}},\ and\ \bibinfo {author} {\bibfnamefont {C.}~\bibnamefont {Monroe}},\
  }\href@noop {} {\bibfield  {journal} {\bibinfo  {journal} {Phys. Rev. Lett.}\
  }\textbf {\bibinfo {volume} {94}},\ \bibinfo {pages} {153602} (\bibinfo
  {year} {2005})}\BibitemShut {NoStop}%
\bibitem [{\citenamefont {Mintert}\ and\ \citenamefont
  {Wunderlich}(2001)}]{mintert2001ion}%
  \BibitemOpen
  \bibfield  {author} {\bibinfo {author} {\bibfnamefont {F.}~\bibnamefont
  {Mintert}}\ and\ \bibinfo {author} {\bibfnamefont {C.}~\bibnamefont
  {Wunderlich}},\ }\href@noop {} {\bibfield  {journal} {\bibinfo  {journal}
  {Physical Review Letters}\ }\textbf {\bibinfo {volume} {87}},\ \bibinfo
  {pages} {257904} (\bibinfo {year} {2001})}\BibitemShut {NoStop}%
\bibitem [{\citenamefont {Harty}\ \emph {et~al.}(2016)\citenamefont {Harty},
  \citenamefont {Sepiol}, \citenamefont {Allcock}, \citenamefont {Ballance},
  \citenamefont {Tarlton},\ and\ \citenamefont {Lucas}}]{harty2016high}%
  \BibitemOpen
  \bibfield  {author} {\bibinfo {author} {\bibfnamefont {T.}~\bibnamefont
  {Harty}}, \bibinfo {author} {\bibfnamefont {M.}~\bibnamefont {Sepiol}},
  \bibinfo {author} {\bibfnamefont {D.}~\bibnamefont {Allcock}}, \bibinfo
  {author} {\bibfnamefont {C.}~\bibnamefont {Ballance}}, \bibinfo {author}
  {\bibfnamefont {J.}~\bibnamefont {Tarlton}},\ and\ \bibinfo {author}
  {\bibfnamefont {D.}~\bibnamefont {Lucas}},\ }\href@noop {} {\bibfield
  {journal} {\bibinfo  {journal} {Phys. Rev. Lett.}\ }\textbf {\bibinfo
  {volume} {117}},\ \bibinfo {pages} {140501} (\bibinfo {year}
  {2016})}\BibitemShut {NoStop}%
\bibitem [{\citenamefont {Srinivas}\ \emph {et~al.}(2021)\citenamefont
  {Srinivas}, \citenamefont {Burd}, \citenamefont {Knaack}, \citenamefont
  {Sutherland}, \citenamefont {Kwiatkowski}, \citenamefont {Glancy},
  \citenamefont {Knill}, \citenamefont {Wineland}, \citenamefont {Leibfried},
  \citenamefont {Wilson} \emph {et~al.}}]{srinivas2021high}%
  \BibitemOpen
  \bibfield  {author} {\bibinfo {author} {\bibfnamefont {R.}~\bibnamefont
  {Srinivas}}, \bibinfo {author} {\bibfnamefont {S.}~\bibnamefont {Burd}},
  \bibinfo {author} {\bibfnamefont {H.}~\bibnamefont {Knaack}}, \bibinfo
  {author} {\bibfnamefont {R.}~\bibnamefont {Sutherland}}, \bibinfo {author}
  {\bibfnamefont {A.}~\bibnamefont {Kwiatkowski}}, \bibinfo {author}
  {\bibfnamefont {S.}~\bibnamefont {Glancy}}, \bibinfo {author} {\bibfnamefont
  {E.}~\bibnamefont {Knill}}, \bibinfo {author} {\bibfnamefont
  {D.}~\bibnamefont {Wineland}}, \bibinfo {author} {\bibfnamefont
  {D.}~\bibnamefont {Leibfried}}, \bibinfo {author} {\bibfnamefont {A.~C.}\
  \bibnamefont {Wilson}}, \emph {et~al.},\ }\href@noop {} {\bibfield  {journal}
  {\bibinfo  {journal} {Nature}\ }\textbf {\bibinfo {volume} {597}},\ \bibinfo
  {pages} {209} (\bibinfo {year} {2021})}\BibitemShut {NoStop}%
\bibitem [{\citenamefont {Wei}\ and\ \citenamefont
  {Norman}(1963)}]{wei1963lie}%
  \BibitemOpen
  \bibfield  {author} {\bibinfo {author} {\bibfnamefont {J.}~\bibnamefont
  {Wei}}\ and\ \bibinfo {author} {\bibfnamefont {E.}~\bibnamefont {Norman}},\
  }\href@noop {} {\bibfield  {journal} {\bibinfo  {journal} {J. Math. Phys.}\
  }\textbf {\bibinfo {volume} {4}},\ \bibinfo {pages} {575} (\bibinfo {year}
  {1963})}\BibitemShut {NoStop}%
\bibitem [{SI()}]{SI}%
  \BibitemOpen
  \href@noop {} {}\bibinfo {note} {{{See Supplemental Material at [URL inserted
  by publisher] for examples of several effective Hamiltonians and a four body
  gate generated by simultaneous displacement and squeezing
  operations.}}}\BibitemShut {Stop}%
\bibitem [{\citenamefont {Egan}\ \emph {et~al.}(2021)\citenamefont {Egan},
  \citenamefont {Debroy}, \citenamefont {Noel}, \citenamefont {Risinger},
  \citenamefont {Zhu}, \citenamefont {Biswas}, \citenamefont {Newman},
  \citenamefont {Li}, \citenamefont {Brown}, \citenamefont {Cetina} \emph
  {et~al.}}]{egan2021fault}%
  \BibitemOpen
  \bibfield  {author} {\bibinfo {author} {\bibfnamefont {L.}~\bibnamefont
  {Egan}}, \bibinfo {author} {\bibfnamefont {D.~M.}\ \bibnamefont {Debroy}},
  \bibinfo {author} {\bibfnamefont {C.}~\bibnamefont {Noel}}, \bibinfo {author}
  {\bibfnamefont {A.}~\bibnamefont {Risinger}}, \bibinfo {author}
  {\bibfnamefont {D.}~\bibnamefont {Zhu}}, \bibinfo {author} {\bibfnamefont
  {D.}~\bibnamefont {Biswas}}, \bibinfo {author} {\bibfnamefont
  {M.}~\bibnamefont {Newman}}, \bibinfo {author} {\bibfnamefont
  {M.}~\bibnamefont {Li}}, \bibinfo {author} {\bibfnamefont {K.~R.}\
  \bibnamefont {Brown}}, \bibinfo {author} {\bibfnamefont {M.}~\bibnamefont
  {Cetina}}, \emph {et~al.},\ }\href@noop {} {\bibfield  {journal} {\bibinfo
  {journal} {Nature}\ ,\ \bibinfo {pages} {1}} (\bibinfo {year}
  {2021})}\BibitemShut {NoStop}%
\bibitem [{\citenamefont {Wang}\ \emph {et~al.}(2020)\citenamefont {Wang},
  \citenamefont {Crain}, \citenamefont {Fang}, \citenamefont {Zhang},
  \citenamefont {Huang}, \citenamefont {Liang}, \citenamefont {Leung},
  \citenamefont {Brown},\ and\ \citenamefont {Kim}}]{wang2020high}%
  \BibitemOpen
  \bibfield  {author} {\bibinfo {author} {\bibfnamefont {Y.}~\bibnamefont
  {Wang}}, \bibinfo {author} {\bibfnamefont {S.}~\bibnamefont {Crain}},
  \bibinfo {author} {\bibfnamefont {C.}~\bibnamefont {Fang}}, \bibinfo {author}
  {\bibfnamefont {B.}~\bibnamefont {Zhang}}, \bibinfo {author} {\bibfnamefont
  {S.}~\bibnamefont {Huang}}, \bibinfo {author} {\bibfnamefont
  {Q.}~\bibnamefont {Liang}}, \bibinfo {author} {\bibfnamefont {P.~H.}\
  \bibnamefont {Leung}}, \bibinfo {author} {\bibfnamefont {K.~R.}\ \bibnamefont
  {Brown}},\ and\ \bibinfo {author} {\bibfnamefont {J.}~\bibnamefont {Kim}},\
  }\href@noop {} {\bibfield  {journal} {\bibinfo  {journal} {Phys. Rev. Lett.}\
  }\textbf {\bibinfo {volume} {125}},\ \bibinfo {pages} {150505} (\bibinfo
  {year} {2020})}\BibitemShut {NoStop}%
\bibitem [{\citenamefont {Ballance}\ \emph {et~al.}(2016)\citenamefont
  {Ballance}, \citenamefont {Harty}, \citenamefont {Linke}, \citenamefont
  {Sepiol},\ and\ \citenamefont {Lucas}}]{ballance2016high}%
  \BibitemOpen
  \bibfield  {author} {\bibinfo {author} {\bibfnamefont {C.}~\bibnamefont
  {Ballance}}, \bibinfo {author} {\bibfnamefont {T.}~\bibnamefont {Harty}},
  \bibinfo {author} {\bibfnamefont {N.}~\bibnamefont {Linke}}, \bibinfo
  {author} {\bibfnamefont {M.}~\bibnamefont {Sepiol}},\ and\ \bibinfo {author}
  {\bibfnamefont {D.}~\bibnamefont {Lucas}},\ }\href@noop {} {\bibfield
  {journal} {\bibinfo  {journal} {Physical review letters}\ }\textbf {\bibinfo
  {volume} {117}},\ \bibinfo {pages} {060504} (\bibinfo {year}
  {2016})}\BibitemShut {NoStop}%
\bibitem [{\citenamefont {Gaebler}\ \emph {et~al.}(2016)\citenamefont
  {Gaebler}, \citenamefont {Tan}, \citenamefont {Lin}, \citenamefont {Wan},
  \citenamefont {Bowler}, \citenamefont {Keith}, \citenamefont {Glancy},
  \citenamefont {Coakley}, \citenamefont {Knill}, \citenamefont {Leibfried}
  \emph {et~al.}}]{gaebler2016high}%
  \BibitemOpen
  \bibfield  {author} {\bibinfo {author} {\bibfnamefont {J.~P.}\ \bibnamefont
  {Gaebler}}, \bibinfo {author} {\bibfnamefont {T.~R.}\ \bibnamefont {Tan}},
  \bibinfo {author} {\bibfnamefont {Y.}~\bibnamefont {Lin}}, \bibinfo {author}
  {\bibfnamefont {Y.}~\bibnamefont {Wan}}, \bibinfo {author} {\bibfnamefont
  {R.}~\bibnamefont {Bowler}}, \bibinfo {author} {\bibfnamefont {A.~C.}\
  \bibnamefont {Keith}}, \bibinfo {author} {\bibfnamefont {S.}~\bibnamefont
  {Glancy}}, \bibinfo {author} {\bibfnamefont {K.}~\bibnamefont {Coakley}},
  \bibinfo {author} {\bibfnamefont {E.}~\bibnamefont {Knill}}, \bibinfo
  {author} {\bibfnamefont {D.}~\bibnamefont {Leibfried}}, \emph {et~al.},\
  }\href@noop {} {\bibfield  {journal} {\bibinfo  {journal} {Physical review
  letters}\ }\textbf {\bibinfo {volume} {117}},\ \bibinfo {pages} {060505}
  (\bibinfo {year} {2016})}\BibitemShut {NoStop}%
\bibitem [{\citenamefont {Magnus}(1954)}]{Magnus1954}%
  \BibitemOpen
  \bibfield  {author} {\bibinfo {author} {\bibfnamefont {W.}~\bibnamefont
  {Magnus}},\ }\href {https://doi.org/https://doi.org/10.1002/cpa.3160070404}
  {\bibfield  {journal} {\bibinfo  {journal} {Communications on Pure and
  Applied Mathematics}\ }\textbf {\bibinfo {volume} {7}},\ \bibinfo {pages}
  {649} (\bibinfo {year} {1954})}\BibitemShut {NoStop}%
\bibitem [{\citenamefont {Horodecki}\ \emph {et~al.}(1999)\citenamefont
  {Horodecki}, \citenamefont {Horodecki},\ and\ \citenamefont
  {Horodecki}}]{horodecki1999general}%
  \BibitemOpen
  \bibfield  {author} {\bibinfo {author} {\bibfnamefont {M.}~\bibnamefont
  {Horodecki}}, \bibinfo {author} {\bibfnamefont {P.}~\bibnamefont
  {Horodecki}},\ and\ \bibinfo {author} {\bibfnamefont {R.}~\bibnamefont
  {Horodecki}},\ }\href@noop {} {\bibfield  {journal} {\bibinfo  {journal}
  {Phys. Rev. A}\ }\textbf {\bibinfo {volume} {60}},\ \bibinfo {pages} {1888}
  (\bibinfo {year} {1999})}\BibitemShut {NoStop}%
\end{thebibliography}%

\end{document}